\documentclass[twocolumn,aps,prd,showpacs,superscriptaddress,groupedaddres]{revtex4}
\usepackage[latin1]{inputenc}
\usepackage{graphicx}
\usepackage{amsmath}
\usepackage{amsfonts}
\usepackage{amssymb}
\usepackage{color}

\newcommand {\lleq} {\lower0.9ex\hbox{ $\buildrel < \over \sim$} ~}
\newcommand{\ggeq} {\lower0.9ex\hbox{ $\buildrel > \over \sim$} ~}

\newcommand{\beq}{\begin{equation}}
\newcommand{\eeq}{\end{equation}}
\newcommand{\ber}{\begin{eqnarray}}
\newcommand{\eer}{\end{eqnarray}}

\newcommand{\sq}{\lower.25ex\hbox{\large$\Box$}}

\begin{document}
\input{epsf.sty}

\title{Dynamics of tachyon dark energy on large scales and its imprint on observed galaxy power spectrum}

\author{Ajay Bassi}
\email{ajay@ctp-jamia.res.in}
\affiliation{Centre for Theoretical Physics, Jamia Millia Islamia, New Delhi - 110025, India.}

\author{Ankan Mukherjee}
\email{ankan.ju@gmail.com}
\affiliation{Department of Physics, Bangabasi College, Kolkata 700009, India.}
\affiliation{Centre for Theoretical Physics, Jamia Millia Islamia, New Delhi - 110025, India.}

\author{Anjan A. Sen}
\email{anjan.sen@ahduni.edu.in,aasen@jmi.ac.in}
\affiliation{School of Arts and Sciences, Ahmedabad University, Ahmedabad 380009, India}
\affiliation{Centre for Theoretical Physics, Jamia Millia Islamia, New Delhi - 110025, India.}

\begin{abstract}

In the present work, we study the large scale matter power spectrum as well as  the observed galaxy power spectrum for non-canonical tachyon field dark energy model considering the full general relativistic perturbation equations. We form a set of coupled autonomous equations including both the  background and linearly perturbed quantities and obtain their solutions numerically with proper set of initial conditions. We consider different scalar field potentials for our study. Deviations from concordance $\Lambda$CDM model are studied for different relevant quantities. Our study shows that non-canonical tachyon dark energy model produces enhanced gravitational potentials, comoving density contrast as well as linear growth factor for matter perturbations compared to $\Lambda$CDM. It is also observed that for tachyon dark energy models, there is suppression of power on large scales compared to both $\Lambda$CDM model as well as previously studied canonical scalar field models.

\end{abstract}



\maketitle

\section{Introduction}
The observed phenomenon of late time cosmic acceleration \cite{Riess:1998cb,Perlmutter:1998np} has brought a drastic change in our understanding of present universe. The genesis of cosmic acceleration is not yet firmly established. An exotic component dubbed as $dark\ energy$ can be introduced in the energy budget of the universe to produce the desired repulsive gravitational effect. Besides the dark energy cosmology, modified gravity theories are also introduced to explain the cosmic acceleration (for comprehensive review, see \cite{Tsujikawa:2010zza}).  

Among different theoretical prescriptions to explain late time cosmic acceleration, the dark energy cosmology is found to be the most consistent with astronomical observations. But we hardly have the knowledge about the actual physical entity of dark energy. Cosmological constant or vacuum energy density is a potential candidate of dark energy \cite{Carroll:2000fy,Peebles:2002gy,Padmanabhan:2002ji}. The cosmological constant ($\Lambda$) along with cold dark matter (CDM) is known as the concordance $\Lambda$CDM model. Though $\Lambda$CDM model is consistent with most of the cosmological observations \cite{Aghanim:2018eyx}, there are certain theoretical issues like the fine tuning problem as well as the cosmic coincidence problem. Apart from this, some of the recent astronomical observations, mainly the local measurement of Hubble constant ($H_0$) \cite{Riess:2019cxk} and the direct measurements of the fluctuations in the matter density distribution in the universe ($S_8$) by KiDS+VIKING-450+DES-Y1 \cite{Asgari:2019fkq} are in tension  with the Planck-$\Lambda$CDM estimation of those parameters.
For all these reasons, time evolving dark energy models are also well emphasized in the literature. In case of time evolving dark energy, the potential candidates are different canonical and non-canonical scalar fields \cite{Ratra:1987rm,Tsujikawa:2013fta,Scherrer:2004au,Padmanabhan:2002cp} or some exotic fluid with specific equation of state \cite{Bento:2003dj}. For a comprehensive review of different time evolving dark energy models, we refer \cite{Copeland:2006wr}. 

In the present  work, we study the evolution of cosmological perturbations in tachyon dark energy model. Tachyon is a non-canonical description scalar field dark energy. Tachyon scalar field was invoked in context of dark energy by Padmanabhan \cite{Padmanabhan:2002cp}. Many more discussions on tachyon dark energy are there in literature \cite{Copeland:2004hq,Bagla:2002yn,Abramo:2003cp,Aguirregabiria:2004xd,Guo:2004dta,Martins:2016lgr}. Spherical collapse of matter overdensity in tachyon dark energy is studied by  Rajvanshi and Bagla \cite{Rajvanshi:2020das} and by Setare, Felegary and Darabi \cite{Setare:2017}. Effects of inhomogeneous tachyon dark energy on cosmological perturbations are studied by Singh, Jassal and Sharma \cite{Singh:2019bfd}.

It is an important task in dark energy cosmology to distinguish the time-varying dark energy model  from the cosmological constant. The possible way to accomplish this is to study the background expansion as well as the evolution of cosmological perturbations. Cosmological perturbations in the matter field and its evolution can be studied from the temperature and polarization spectrum of cosmic microwave background (CMB) \cite{Aghanim:2018eyx} and also from the observed galaxy power spectrum \cite{Tegmark:2003uf}. CMB observation by Planck along with other observational data have ensured unprecedented constraints on cosmological parameters \cite{Aghanim:2018eyx}. But most of these observations probe sub-horizon scale physics where the Newtonian approximations for cosmological perturbation is valid and the dark energy perturbations can be safely ignored. Thus all the dark energy parameters, constrained in Planck observations, are related to the background evolution of dark energy. Future observations like the LSSR (Large Synoptic Survey Telescope) \cite{Foley:2018}, SKA (Squre Kilometer Array) \cite{Maartens:2015mra} will provide a wide redshift range sky survey in optical and radio observations and a much more sophisticated map of the distribution of matter in the universe. These type of observations will be highly effective to study the general relativistic (GR) effects in the evolution of cosmological perturbations where the inhomogeneities in dark energy could not be ignored. As the cosmological constant ($\Lambda$) is homogeneous, these future observations would be the smoking gun to distinguish the cosmological constant from  the time varying dark energy models and would also be effective to check the viability of various dark energy models.
 
As already mentioned, the present analysis is carried out for tachyon dark energy. The full general relativistic effects on the evolution of linear perturbations and galaxy power spectrum are studied. We formulate a set of autonomous system of equations which are studied numerically with proper initial conditions. Quintessence scalar field dark energy perturbation and its scale dependence was studied by Unnikrishnan, Jassal and Seshadri \cite{Unnikrishnan:2008qe}. In case of tracker quintessence, the galaxy power spectrum incorporating GR corrections and its imprint on the neutral hydorgen distribution in the universe are studied by Duniya, Bertacca and Maartens \cite{Duniya:2013eta}. Dinda and Sen have studied the galaxy power spectrum in inhomogeneous thawing scalar field dark energy \cite{Dinda:2016ibo}. Recently, Singh, Jassal and Sharma \cite{Singh:2019bfd} have studied the terturbations in tachyon dark nergy and its effects on the clustering of dark matter and a comperative study of linear perturbation in quintessence and tachyon is carried out by Rajvanshi {\it et al} \cite{Rajvanshi:2021afc}.

The paper is organized as the following. In section \ref{background}, the background evolution equations for the present models are discussed. The peturbation equations with general relativistic corrections and their solutions are discussed in section \ref{RelativisticPerturbation}. In section \ref{powerspectrum}, the different power spectrum and their deviation from $\Lambda$CDM at different redshift are presented. Finally, in section \ref{conclu}, we have concluded with an overall discussion about the results.

\section{Background Evolution}
\label{background}

A non-canonical description of scalar field dark energy, namely the tachyon, is studied in the present work. 
We consider Dirac-Born-Infeld(DBI) type of action to study the dynamics of the tachyon scalar field 
\begin{equation}
\mathcal{S}=\int {
-V(\phi)\sqrt{1-\partial^\mu\phi\partial_\mu\phi}}\sqrt{-g} d^4x.
\label{eq1}
\end{equation}
Here $V(\phi)$ is the potential for non-canonical scalar field $\phi$. The energy density and pressure of the tachyon scalar field are respectively given by \cite{Bagla:2002yn}
\begin{equation}
\bar{\rho}_{\phi}=\frac{V(\phi)}{\sqrt{1-\dot{\phi}^2}},
\label{eq3}
\end{equation}
\begin{equation}
\bar{P}_{\phi}=-V(\phi)\sqrt{1-\dot{\phi}^2},
\label{eq2}
\end{equation}
where overhead dot represents the derivative w.r.t. cosmic time. From the action, given in equation (\ref{eq1}), the equation of motion for scalar field is obtained as,
\begin{equation}
\ddot{\phi}+3H\dot{\phi}(1-\dot{\phi}^2)+\frac{V_{,\phi}}{V}(1-\dot{\phi}^2)=0
\label{eq4}
\end{equation}
where subscript $,\phi$ is the derivative w.r.t. the scalar field $\phi$. The Hubble parameter ($H$) in a spatially flat FLRW universe is expressed,
\begin{equation}
H^2=\frac{\bar{\rho}_{\phi}+\bar{\rho}_m}{3}
\label{eq5}
\end{equation}
where $\bar{\rho}_{m}$ is the energy density of the background matter which includes the contribution from both the dark matter and baryons. 

\begin{figure}
\includegraphics[width=0.97\linewidth]{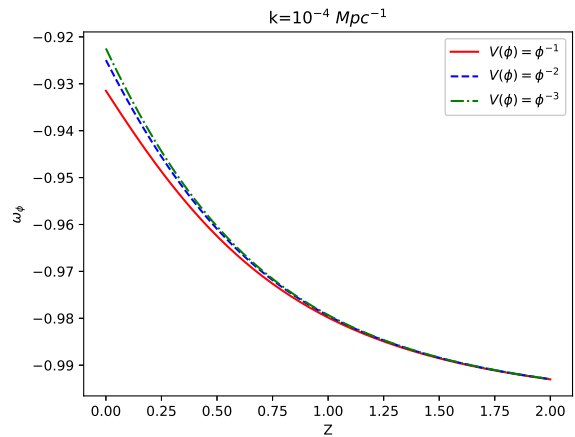}
\caption{Equation of state for the tachyon scalar field $w_{\phi}$ as a function of redshift $z$ at large scale $k=10^{-4} Mpc^{-1}$ and for different potentials with $\Omega_{m0} = 0.28$ and $\lambda_{i} = 0.7$.}
\label{figa}
\end{figure}

\section{Relativistic Perturbation}
\label{RelativisticPerturbation}

We consider conformal Newtonian gauge with vanishing anisotropic stress for the flat FLRW spacetime with perturbed metric

\begin{equation}
ds^{2} = a^{2}(\tau)\left[(1+2\Phi) d\tau^{2}-(1-2\Phi) d\vec{x}.d\vec{x}\right],
\label{eq6}
\end{equation}

\noindent
where $ \tau $ is the conformal time, $a(\tau)$ is the conformal scale factor, $\vec{x}$ are the comoving coordinates and $\Phi$ is the gravitational potential. The linearized Einstein equations obtained for the above perturbed metric (equation (\ref{eq6})) are written as \cite{Dinda:2016ibo},

\begin{equation}
\nabla^{2} \Phi - 3 \mathcal{H} (\Phi' + \mathcal{H} \Phi) = 4 \pi G a^{2} \sum_{i} \delta \rho_{i},
\label{eq7}
\end{equation}

\begin{equation}
\Phi' + \mathcal{H} \Phi = 4 \pi G a^{2} \sum_{i} (\bar{\rho_{i}} + \bar{P_{i}}) v_{i},
\label{eq8}
\end{equation}

\begin{equation}
\Phi'' + 3 \mathcal{H} \Phi' + (2 \mathcal{H}' + \mathcal{H}^{2}) \Phi = 4 \pi G a^{2} \sum_{i} \delta P_{i},
\label{eq9}
\end{equation}

\noindent
where prime denotes the derivative w.r.t. the conformal time $ \tau $, $ \bar{P_{i}} $ and $ \bar{\rho_{i}} $ represent the background pressure and energy density of each component, namely the matter and tachyon filed and $ \mathcal{H} $ denotes the conformal Hubble parameter . $ \delta P_{i} $ , $ \delta \rho_{i} $ and $ v_{i} $ are the linear order perturbed quantities for the background pressure, energy density and velocity field respectively. $ \vec{v_{i}} = - \vec{\nabla} v_{i} $ defines the irrotational component of the velocity field. From equations (\ref{eq7}) and (\ref{eq8}), one gets the relativistic Poisson equation as, 

\begin{equation}
\nabla^{2} \Phi = 4 \pi G a^{2} \sum_{i} \bar{\rho_{i}} \Delta_{i},
\label{eq10}
\end{equation}

\noindent
where $ \Delta_{i} = \delta_{i} + 3 \mathcal{H} (1 + w_{i}) v_{i} $ represents the gauge invariant comoving energy density contrast for the i-th component. $\Delta_{i}$ is the correct tracer for the gravitational potential on large scales. The relativistic continuity and Euler equations can be obtained from the conservation of stress-energy tensor as \cite{Dinda:2016ibo},

\begin{equation}
\delta' + 3 \mathcal{H} (\dfrac{\delta P}{\delta \rho} - \dfrac{\bar{P}}{\bar{\rho}}) \delta = (1 + \dfrac{\bar{P}}{\bar{\rho}}) (\theta + 3 \Phi'),
\label{eq11}
\end{equation}

\noindent
and

\begin{equation}
\theta' + 3 \mathcal{H} (\dfrac{1}{3} - \dfrac{\bar{P}'}{\bar{\rho}'}) \theta = \dfrac{\nabla^{2} \delta P}{\bar{\rho} + \bar{P}} + \nabla^{2} \Phi,
\label{eq12}
\end{equation}

\noindent
respectively, where $ \theta = - \vec{\nabla}.\vec{v} $ and $ \delta=\frac{\delta \rho}{\bar{\rho}}$. Finally the evolution equations of perturbed energy density, pressure and velocity at linear order for tachyon scalar field are given as,

\beq
\delta\rho_\phi=\frac{V(\phi)}{(1-\dot{\phi^2})^{3/2}}(\dot{\phi}\delta\dot{\phi}-\Phi\dot{\phi^2})+\frac{V_\phi\delta\phi}{\sqrt{1-\dot{\phi^2}}},
\label{eq13}
\eeq
\beq
\delta P_\phi=\frac{V(\phi)}{\sqrt{1-\dot{\phi^2}}}(\dot{\phi}\delta\dot{\phi}-\Phi\dot{\phi^2})-V_\phi\delta\phi\sqrt{1-\dot{\phi^2}},
\label{eq14}
\eeq
\beq
a(\bar{\rho_\phi}+\bar{P_\phi})v_\phi=V(\phi)\frac{\dot{\phi}\delta\phi}{\sqrt{1-\dot{\phi^2}}}.
\label{eq15}
\eeq
Next, we define the following dimensionless parameters related to the background  and perturbed quantities for the tachyon field:
\ber
\nonumber
x=\dot \phi, ~~~~~~~~
y= \frac{\sqrt{V(\phi)}}{\sqrt{3}H}, \\ 
\nonumber
\lambda=-\frac{V_{,\phi}}{V^{3/2}},  ~~~~~
\Gamma =V\frac{V_{,\phi\phi}}{(V_{,\phi})^2},\\
\nonumber
\delta\phi=\frac{\dot{\phi}}{H}q, ~~~~
\Omega_\phi=\frac{y^2}{\sqrt{1-x^2}},\\
\gamma_\phi=1+\omega_\phi=\dot{\phi^2}=x^2
\label{eq16}
\eer

The $x$ here is just a dimentionless parameter and is different from the comoving coordinates in equation (\ref{eq6}). $\Omega_{\phi}$ is the density parameter and $w_{\phi}$ is the equation of state parameter for the tachyon scalar field $\phi$. We can now form a set of autonomous system of equations involving the quantities defined in equation (\ref{eq16}) to study the different quantities associated with both the background  and perturbed universe \cite{Scherrer:2007pu},
\ber
\nonumber
\gamma_\phi'=-6\gamma_\phi(1-\gamma_\phi)+2\sqrt{3\gamma_\phi\Omega_\phi}\lambda(1-\gamma_\phi)^{5/4},\\
\nonumber
\Omega_\phi'=3\Omega_\phi(1-\gamma_\phi)(1-\Omega_\phi),~~~~~~~~~~~~~~~~~~~~~~~\\
\nonumber
\lambda'=-\sqrt{3\gamma_\phi\Omega_\phi}\lambda^2(1-\gamma_\phi)^{1/4}(\Gamma-{3/2}),~~~~~~~~\\
\nonumber
\mathcal{H'}=-\frac{1}{2}(1+3\Omega_\phi(\gamma_\phi-1))\mathcal{H},~~~~~~~~~~~~~~~~~~~~\\
\nonumber
\Phi'= \Phi_1,~~~~~~~~~~~~~~~~~~~~~~~~~~~~~~~~~~~~~~~~~~~~~~~\\
\nonumber
q'= q_1,~~~~~~~~~~~~~~~~~~~~~~~~~~~~~~~~~~~~~~~~~~~~~~~~\\
\nonumber
\Phi_1'=-(1+B)\Phi_1 -(2B-3+\frac{3}{2}\Omega_\phi\gamma_\phi)\Phi~~~~~\\
\nonumber
+\frac{3}{2}\Omega_\phi\gamma_\phi[q_1 +q(3\gamma_\phi-B+g(1-\gamma_\phi))],\\
\nonumber
q_1'=-(g-3\gamma_\phi-B)q_1-B_qq-(3\gamma_\phi-4)\Phi_1\\
+(g-6\gamma_\phi)\Phi.
\label{eq17}
\eer
Here prime represents derivative w.r.t. $N=log(a)$. We have defined $B=1.5(1-(\gamma_\phi-1)\Omega_\phi), g=2\lambda\sqrt{\frac{3\Omega_\phi}{\gamma_\phi}}(1-\gamma_\phi)^{1/4}$ and $B_q=-B'+(g-6\gamma_\phi)(3-B)+\frac{k^2}{\mathcal{H}^2}(1-\gamma_\phi).$

Matter density contrast and peculiar velocity for matter are obtained from the Fourier space solutions of equations (\ref{eq7}), (\ref{eq8}), (\ref{eq13}) and (\ref{eq15}) as,
\ber
\nonumber
\delta_m = -\frac{2}{\Omega_m}[\Phi_1+\Phi(1-\frac{\Omega_\phi\gamma_\phi}{2(1-\gamma_\phi)}+\frac{k^2}{3\mathcal{H}^2})~~~~~~~~~~~~~~~\\
\nonumber
+\frac{\Omega_\phi\gamma_\phi}{2(1-\gamma_\phi)}(q_1+q(3\gamma_\phi-B)],~~~~~~~~~~~~~~~~~~~~~~~~~~\\
y_m = 3\mathcal{H}v_m=\frac{2}{\Omega_m}[\Phi_1+\Phi-1.5q\Omega_\phi\gamma_\phi].~~~~~~~~~~~~~~~~~~
\label{eq18}
\eer
Using equation (\ref{eq18}), we can define the gauge invariant comoving matter density contrast as $\Delta_m=\delta_m+y_m$.

\begin{figure*}
\begin{tabular}{c c c }
\includegraphics[width=0.5\linewidth]{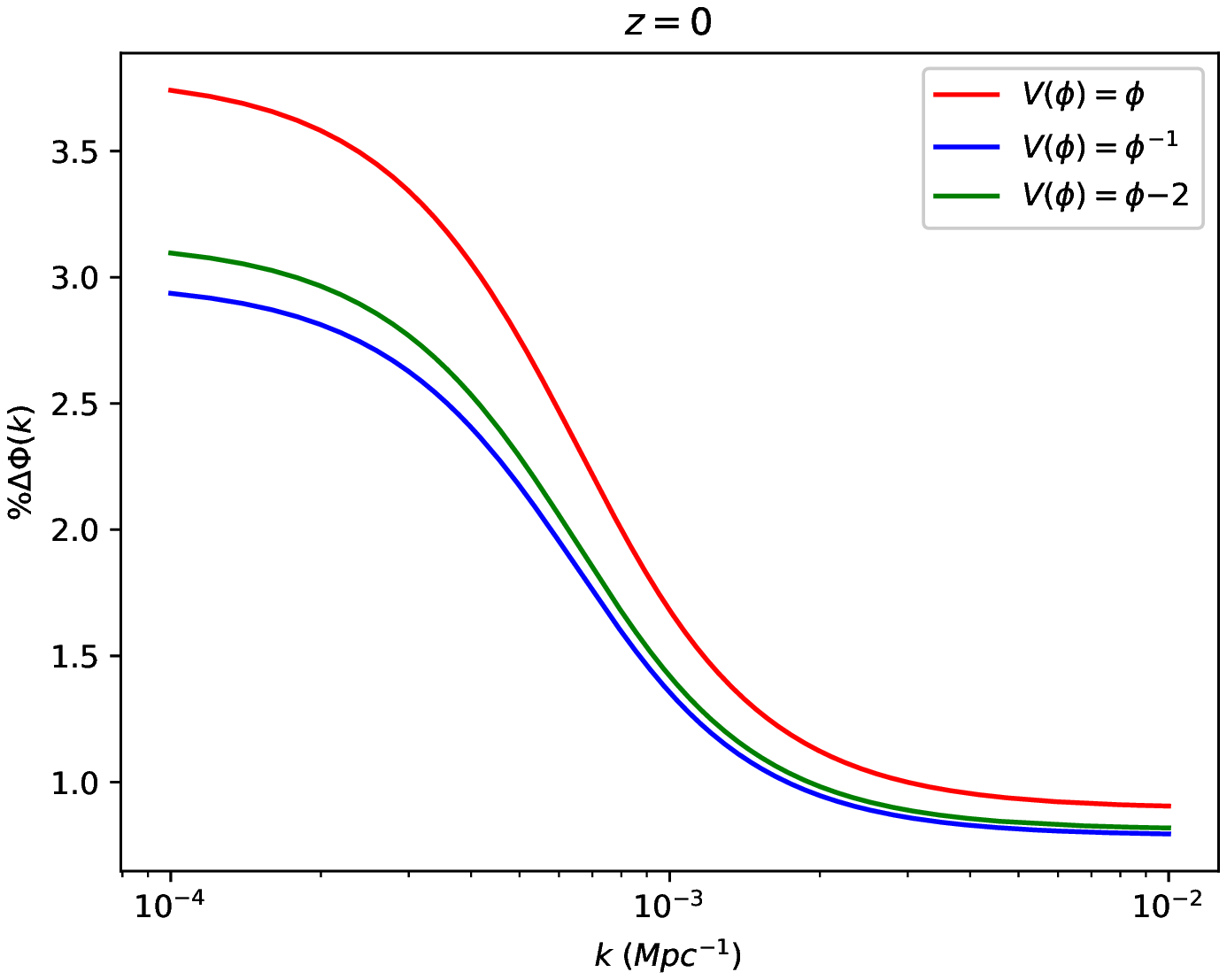}\\
\includegraphics[width=0.5\linewidth]{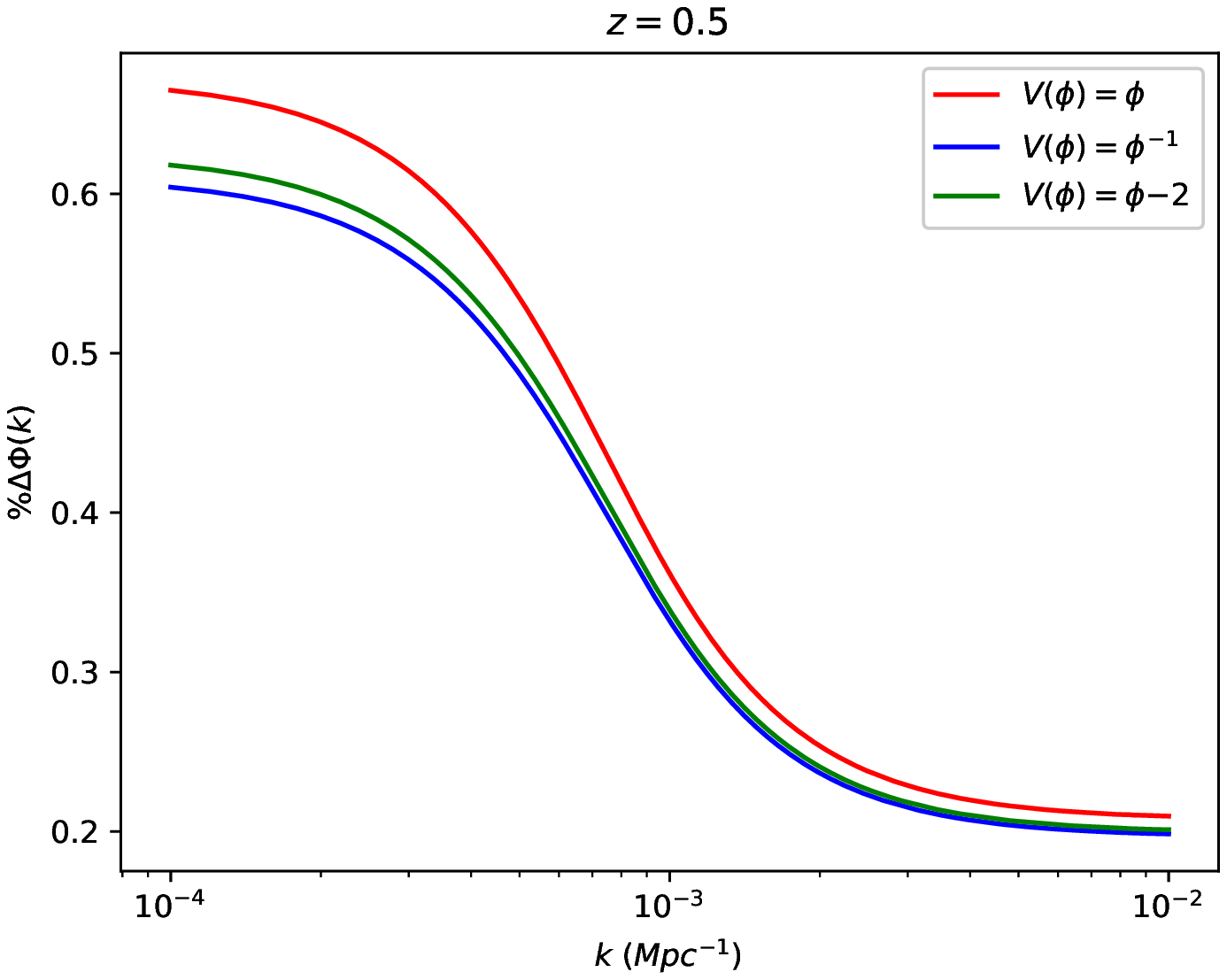}
\includegraphics[width=0.5\linewidth]{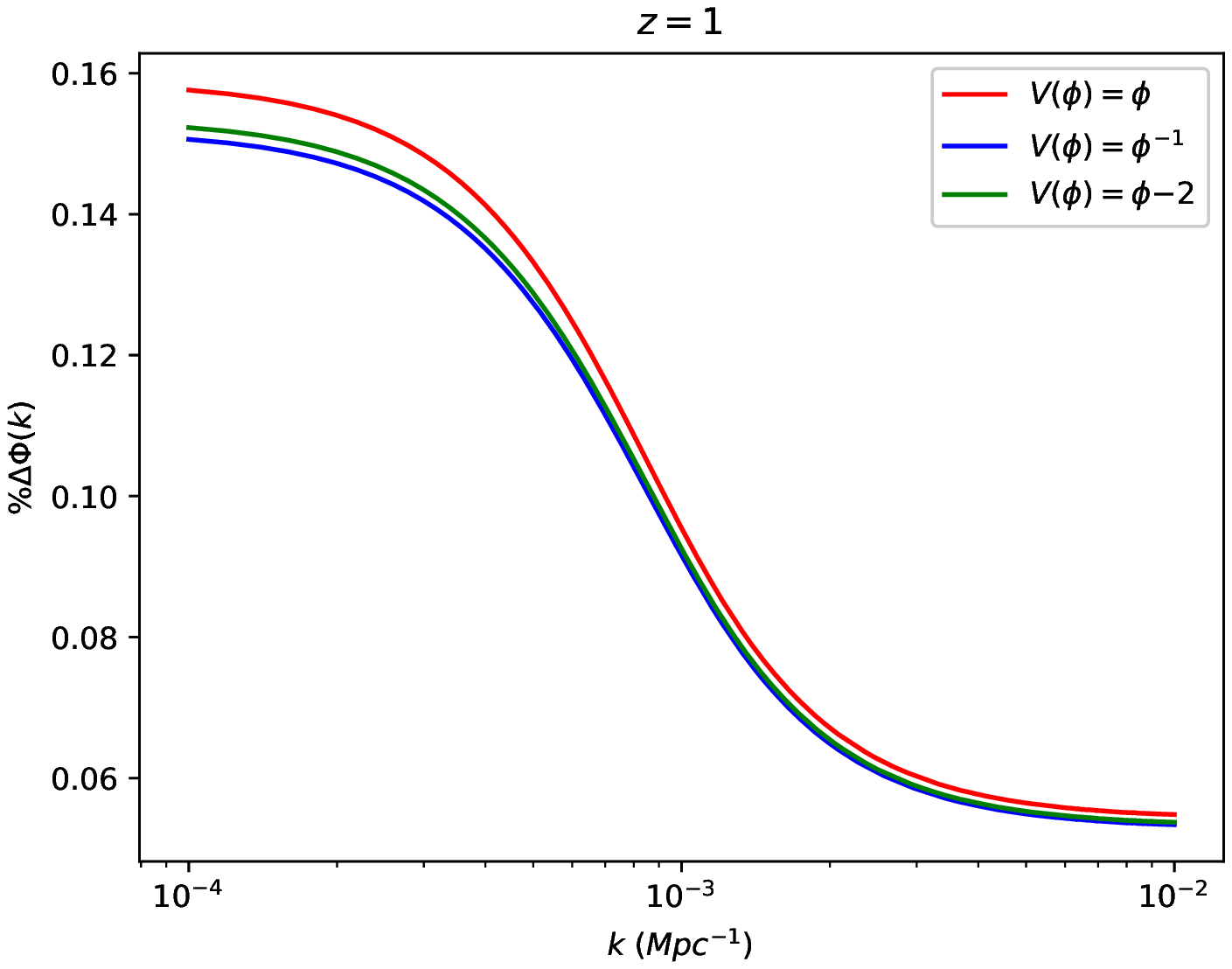}
\end{tabular}
\caption{ Percentage deviation in gravitational potential $\Phi$ from $\Lambda CDM$ model as a function of $k$ with $\Omega_{m0} = 0.28$ and $\lambda_{i} = 0.7$. Here and in subsequent plots, we use $\% \Delta X = (X^{\phi} / X^{\Lambda} - 1)\times 100$.}
\label{figb}
\end{figure*}


\begin{figure*}
\begin{tabular}{c c c }
\includegraphics[width=0.5\linewidth]{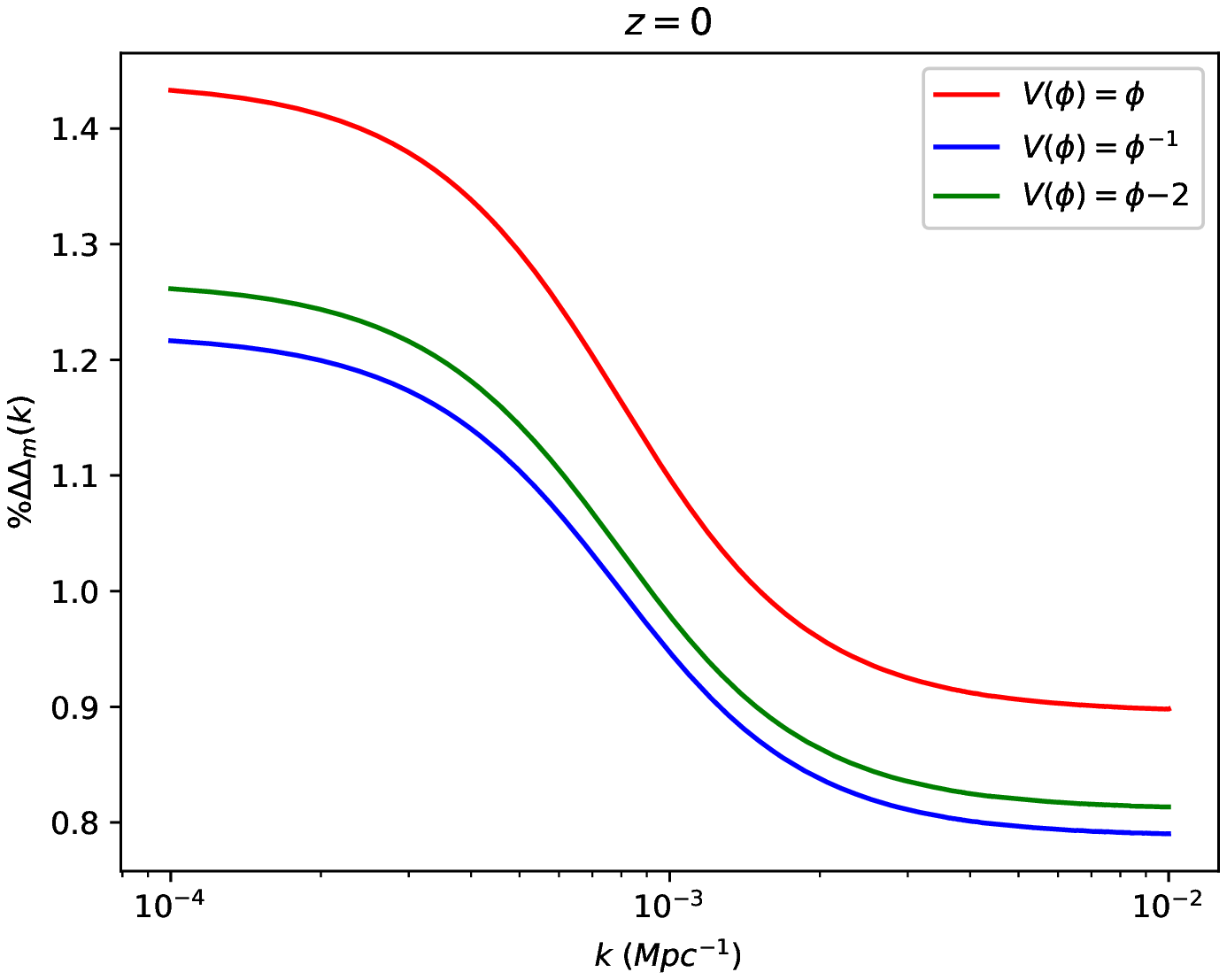}\\
\includegraphics[width=0.5\linewidth]{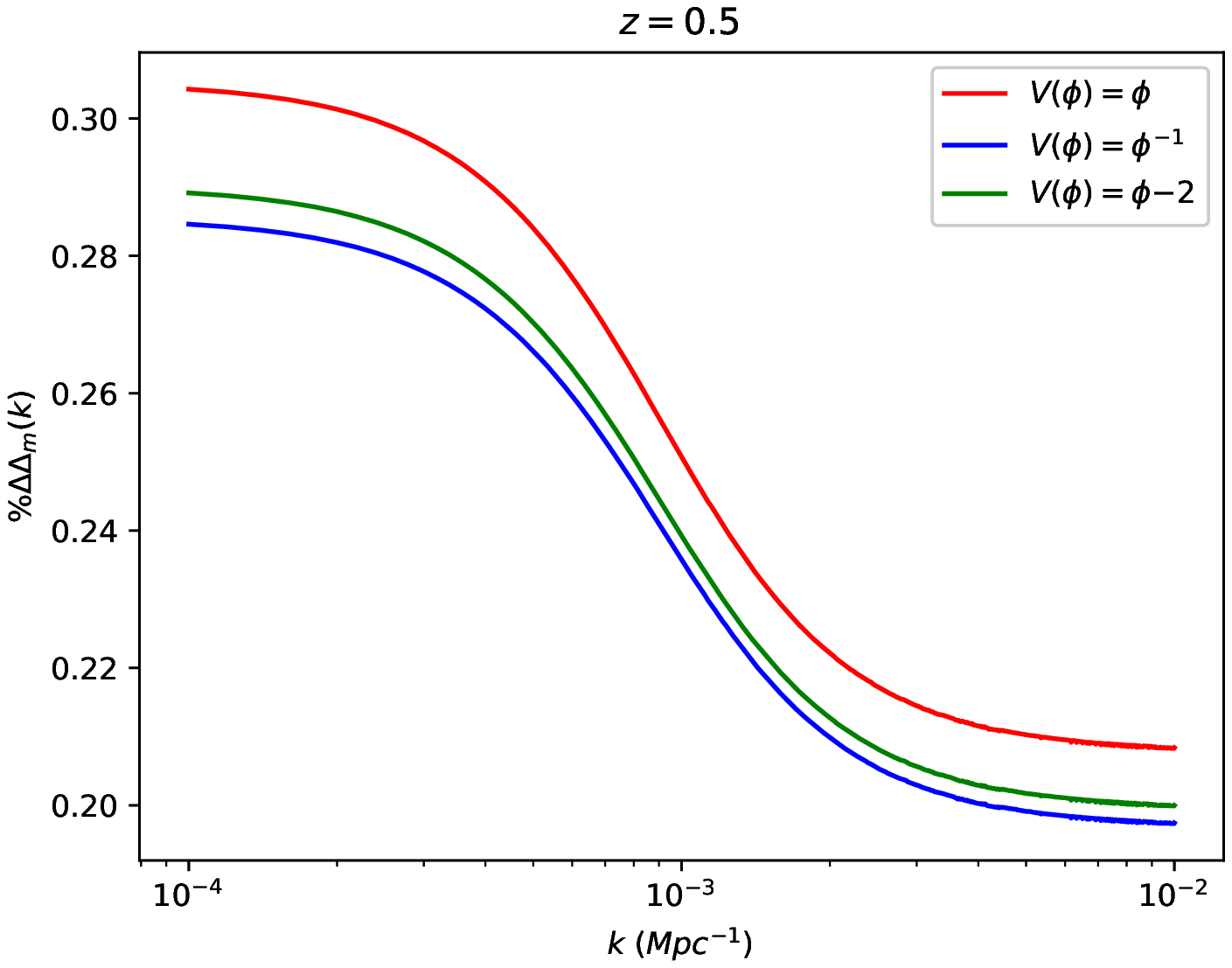}
\includegraphics[width=0.5\linewidth]{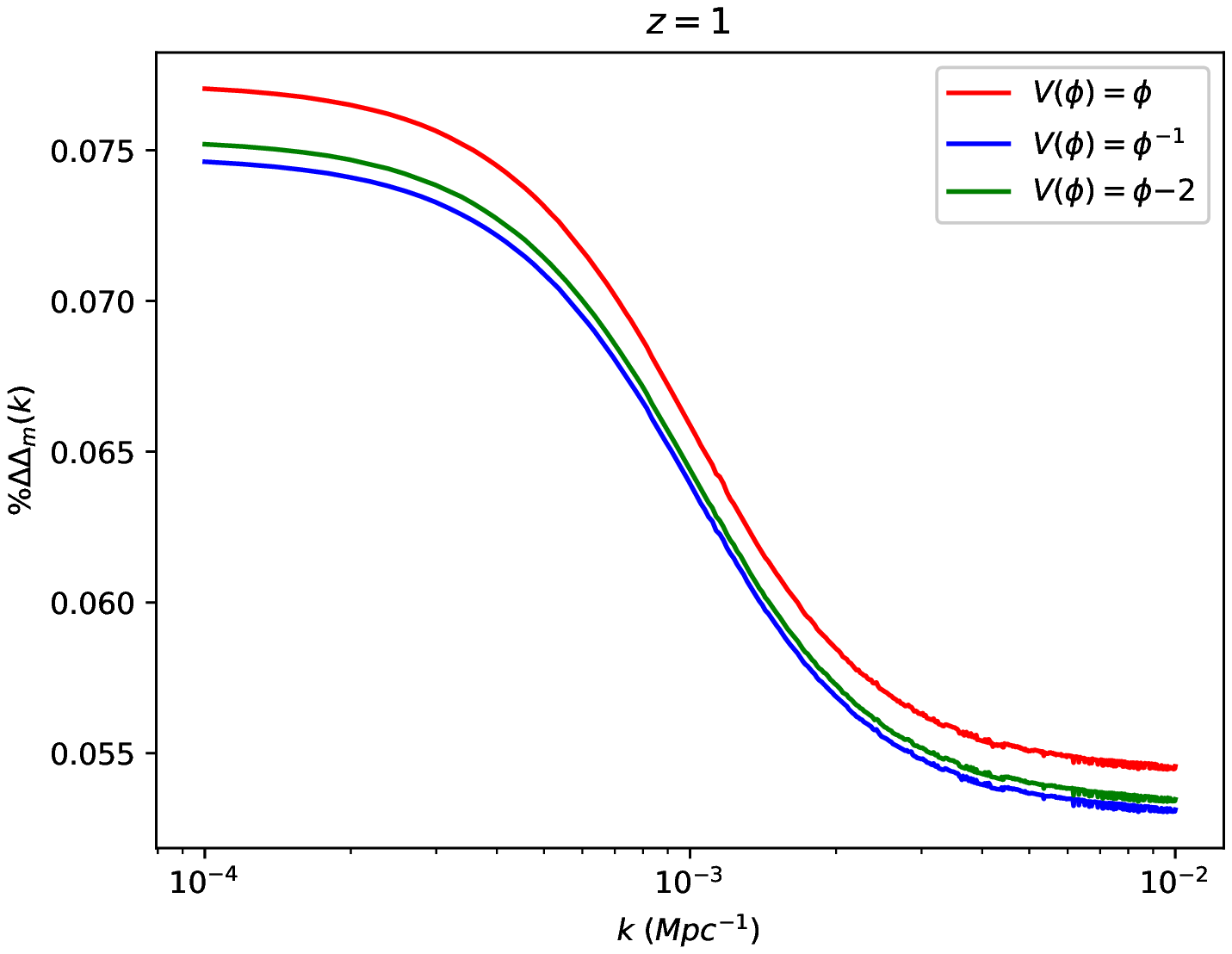}
\end{tabular}
\caption{ Percentage deviation in comoving density contrast $\Delta_m$ from $\Lambda CDM$ model as a function of $k$.}
\label{figc}
\end{figure*}


\begin{figure*}
\begin{tabular}{c c c}
\includegraphics[width=0.5\linewidth]{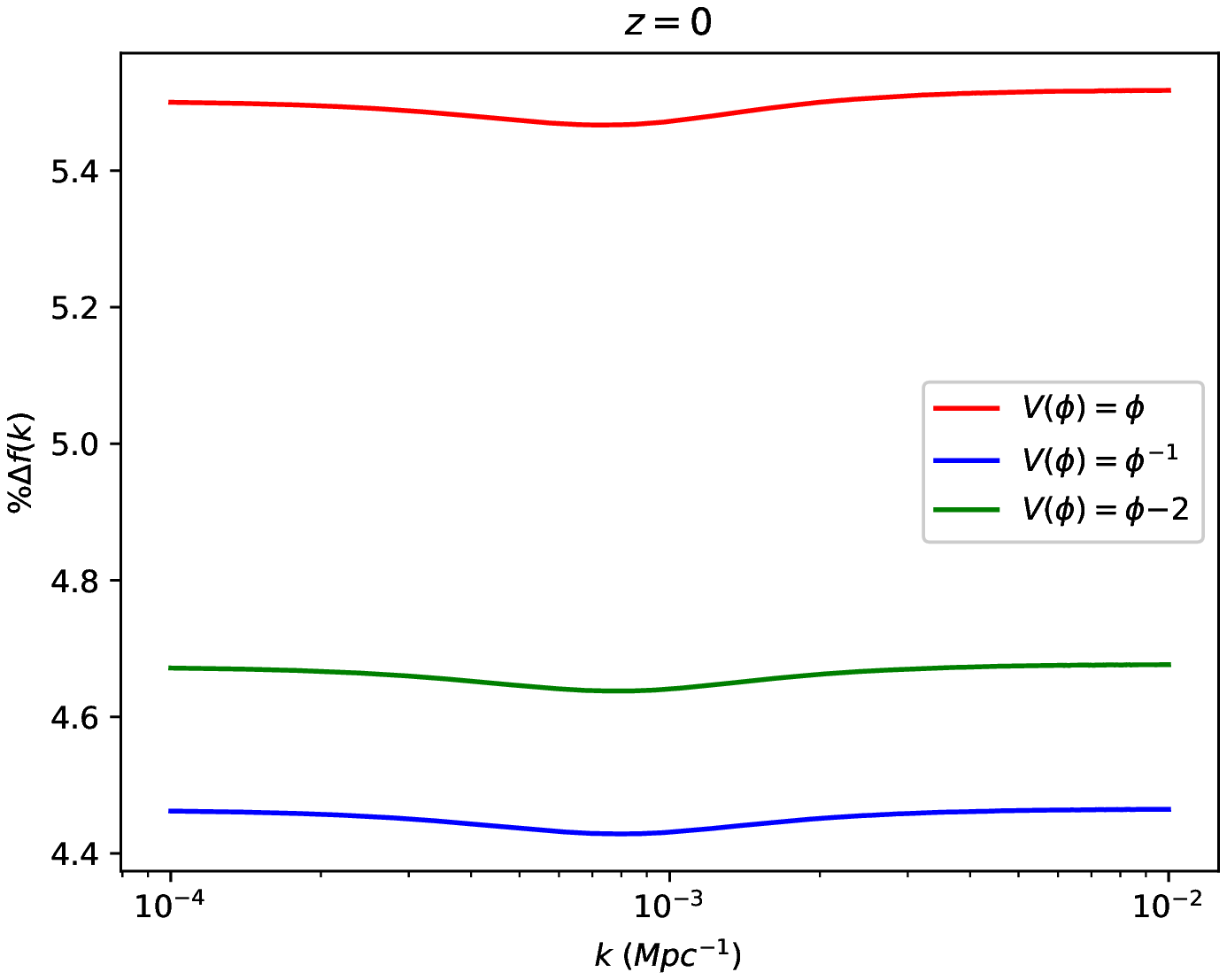}\\
\includegraphics[width=0.5\linewidth]{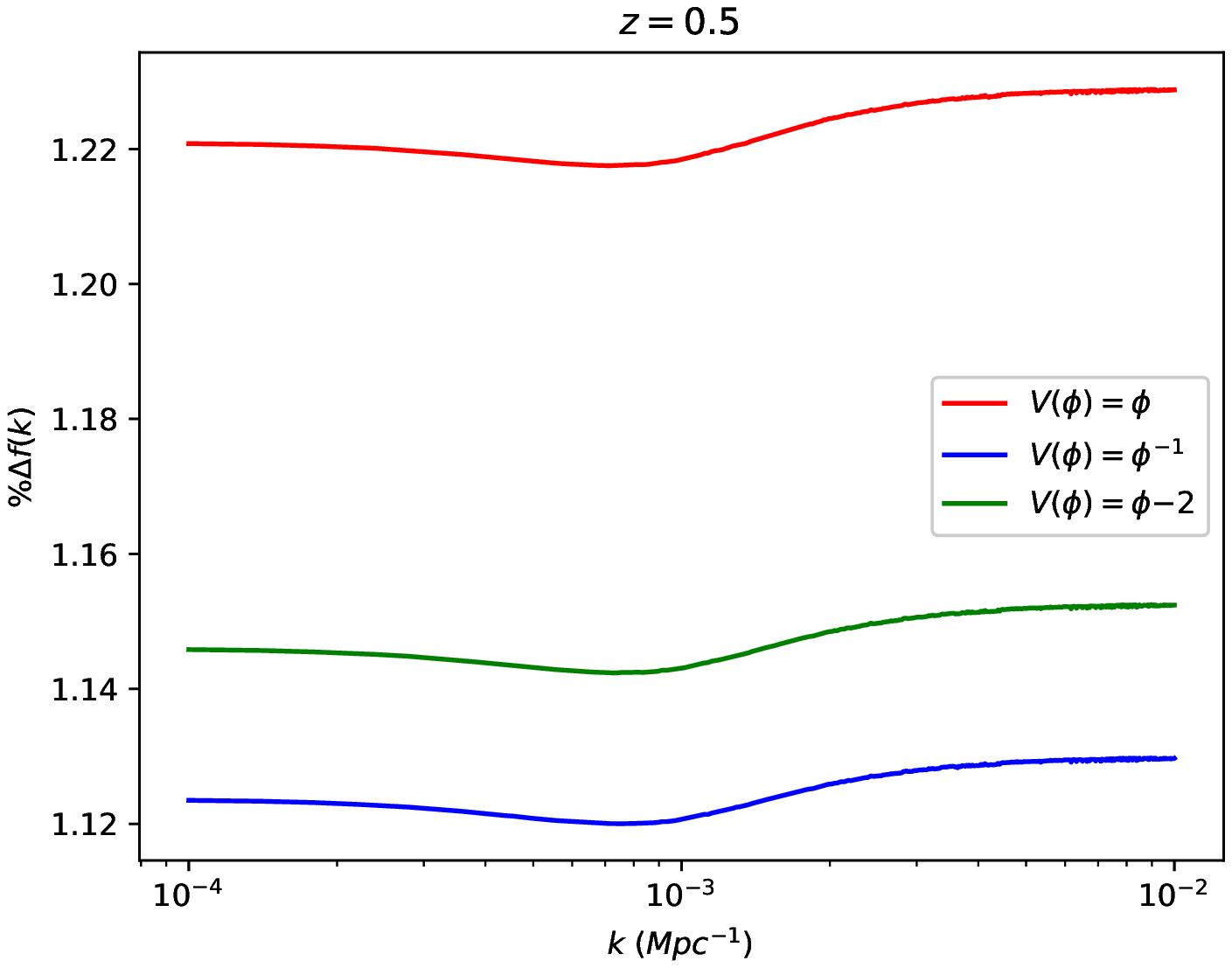}
\includegraphics[width=0.5\linewidth]{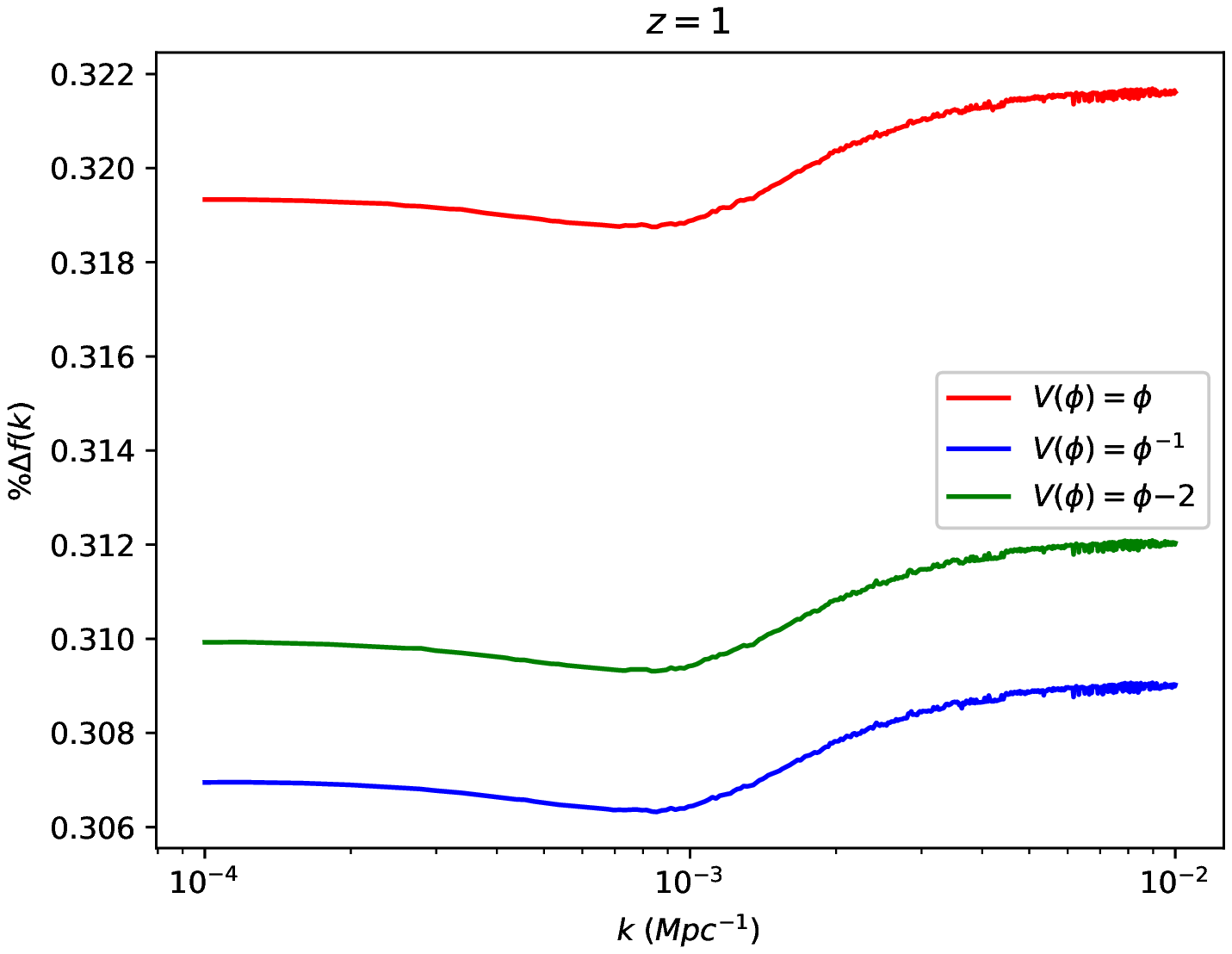}
\end{tabular}
\caption{ Percentage deviation in $f$ from $\Lambda CDM$ model as a function of $k$.}
\label{figd}
\end{figure*}


\subsection{Initial conditions}     
  
One needs to set the initial conditions for ($\gamma, \Omega_{\phi}, \lambda, \mathcal{H}$) for the background universe and  ($\Phi, \Phi', q, q' $) for the perturbed universe to solve the set of autonomous equations defined in equations (\ref{eq17}). We fix the initial conditions at decoupling epoch ($z=1000$), when the universe was matter dominated and contribution from dark energy was negligible. For this, we follow the same procedure as described in \cite{Dinda:2016ibo}. The scalar field is frozen initially at $w_{\phi} \sim -1$ due to large Hubble friction ($3H\dot{\phi}$ term in eq.(\ref{eq4})), such that $\gamma_{i} \sim 0$, but we set it at very small value $\gamma_{i}=10^{-7}$. $\Omega_\phi$ is negligible initially at $z=1000$ because universe was matter dominated. $\lambda$ gives the slope of the potential and determines the evolution of the scalar field. We set $\lambda_{in}<<1$ so that the scalar field remains frozen to the initial value of equation of state $\omega_\phi\sim-1$ and behaves like cosmological constant initially. We fix the initial values of $\Omega_\phi, \lambda$ and $\mathcal{H}$ in a manner so that we get the desired values of $\Omega_{\phi0}$ and $\mathcal{H}_0$ at present redshift $z=0$. 

One can ignore the contribution of dark energy at $z=1000$ as the universe was matter dominated at that redshift and hence we set $q=\dfrac{dq}{dN}=0$ initially. Moreover, the gravitational potential $\Phi$ being constant during matter domination, we set the initial value of gravitational potential using equation (\ref{eq10}) and relation $\Delta_m \sim a$ (during matter domination) as
\beq
\Phi_{in}=-\frac{3}{2}\frac{\mathcal{H}^2_{in}}{k^2}a_{in},
\label{eq19}
\eeq
which is a constant and hence $\dfrac{d\Phi}{dN}=0$ initially.

\subsection{Behaviour of cosmological parameters}

To get the desired results, we fix $\Omega_{m0}=0.28, \lambda_i=0.7$ and $\mathcal{H}_0=70 km/s/Mpc$. These values are consistent with different cosmological observations including CMB by Planck and the overall behaviour of our final results are not sensitive to these values. With the initial conditions set as above, we solve the set of autonomous equations (\ref{eq17}) and study the dynamics of the different cosmological parameters. We are considering power-law potentials, more specifically linear, inverse and inverse-squared potentials.

In figure \ref{figa}, we show the behaviour of equation of state parameter $(\omega_\phi=\gamma_\phi-1)$ as a function of redshift for the different potentials. We set the identical initial conditions for all the potentials and $\omega_\phi$ remains freeze at $\omega_\phi=-1$ initially and thaws away from cosmological constant type behaviour in the near past.

In figure \ref{figb}, we study the behaviour of gravitational potential in comparison to the $\Lambda$CDM case. We show the percentage deviation in the gravitational potential $\Phi$ of the tachyon dark energy from $\Lambda CDM$ model for different types of potentials. For redshifts $z\neq 0$, the deviation is less than $1\%$ for all scales, whereas for $z=0$, the deviation is around $3-4\%$ for large scales and around $1\%$ at small scales. Also the linear potential results the highest deviation compared to other potential. This is similar to the canonical scalar dark energy model \cite{Dinda:2016ibo}. We should stress that the small scale behaviour in tachyon  dark energy model is primarily governed by its background evolution whereas on the large scales, effect of perturbation in tachyon field plays a significant role.

In figure \ref{figc}, we study the behaviour of the gauge invariant matter density contrast $\Delta_m$. The behaviour is similar to the gravitational potential, but with comparatively smaller deviation from $\Lambda$CDM case.

Next, we define the quantity $f$ which depends on the velocity field perturbations and gives rise to the redshift space distortion

\beq
f=-\frac{k^2 v_{m}}{{\mathcal{H}}\Delta_{m}}.
\label{eq20}
\eeq

In figure \ref{figd}, we show the deviatin in $f$ from $\Lambda CDM$ model for different scalar field potentials. For redshift $z=0$, the deviation in $f$ is smaller than $6\%$ and for higher redshifts, the deviation is even smaller for all potentials considered. There is hardly any scale dependency which shows that the contribution to the deviation in $f$ is from background expansion only.

\section{The observed galaxy power spectrum}
\label{powerspectrum}

\begin{figure}
\includegraphics[width=0.96\linewidth]{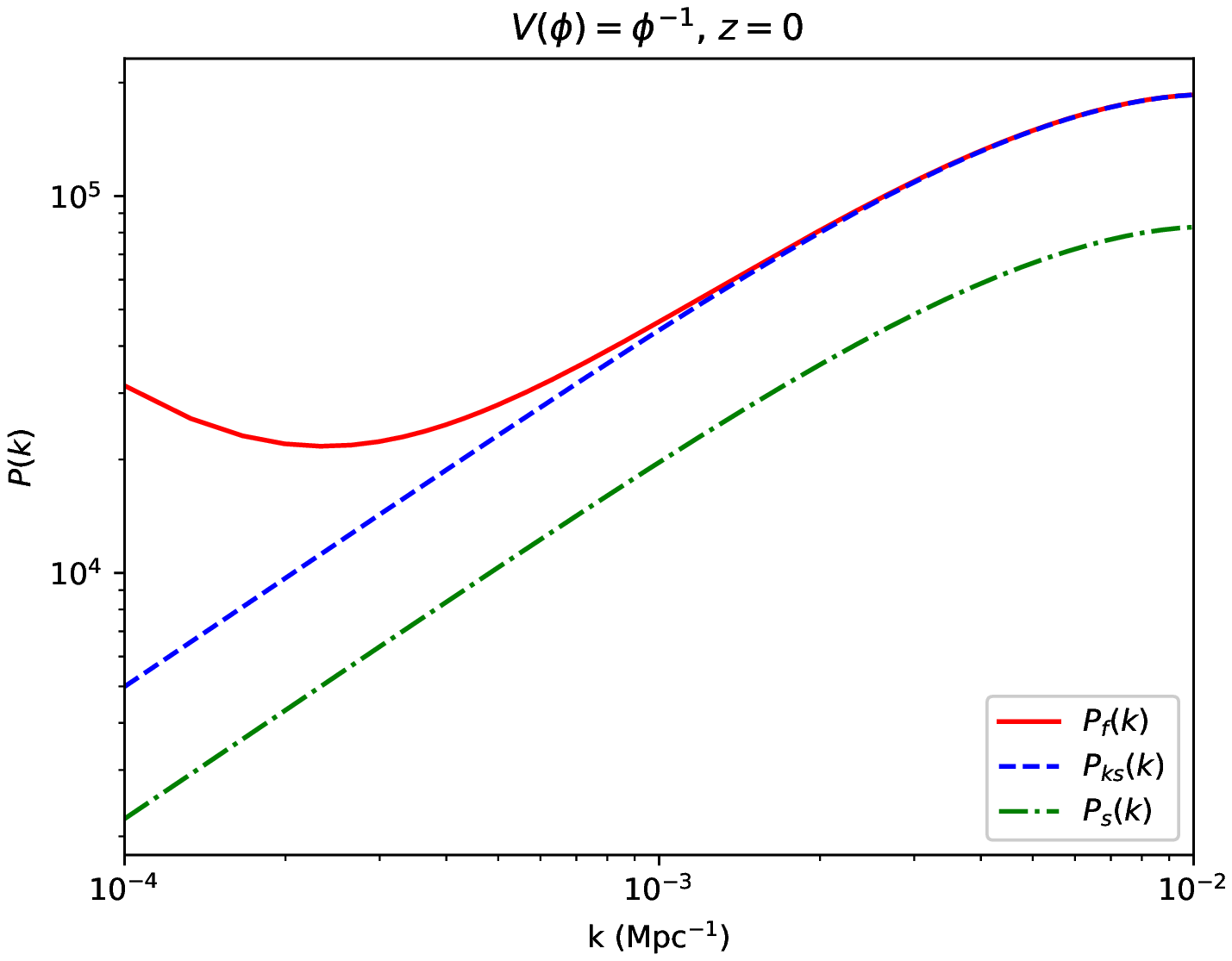}
\caption{Continuous, dashed and dashed-dotted lines for the full observed galaxy power spectrum $P(k)$ given by eqn. (\ref{eq24}), the galaxy power spectrum $P_{ks}(k)$ by taking only the Kaiser redshift term ( first term inside the square bracket in eqn. (\ref{eq24})) and the standard matter power spectrum $P_s(k)$ given by eqn. (\ref{25}) as a function of $k$. }
\label{fige}
\end{figure}

\begin{figure*}
\begin{tabular}{c c c }
\includegraphics[width=0.33\linewidth]{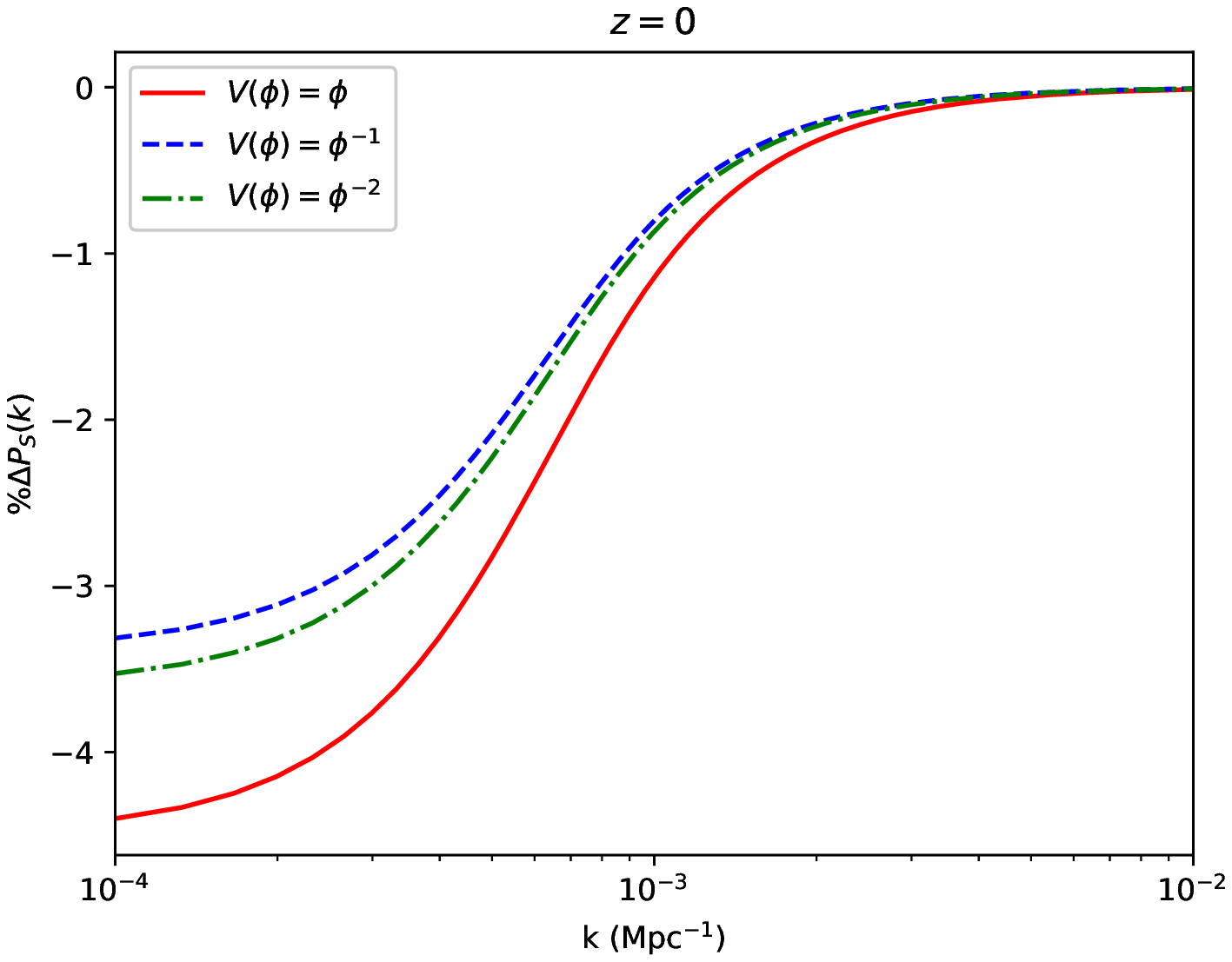}
\includegraphics[width=0.33\linewidth]{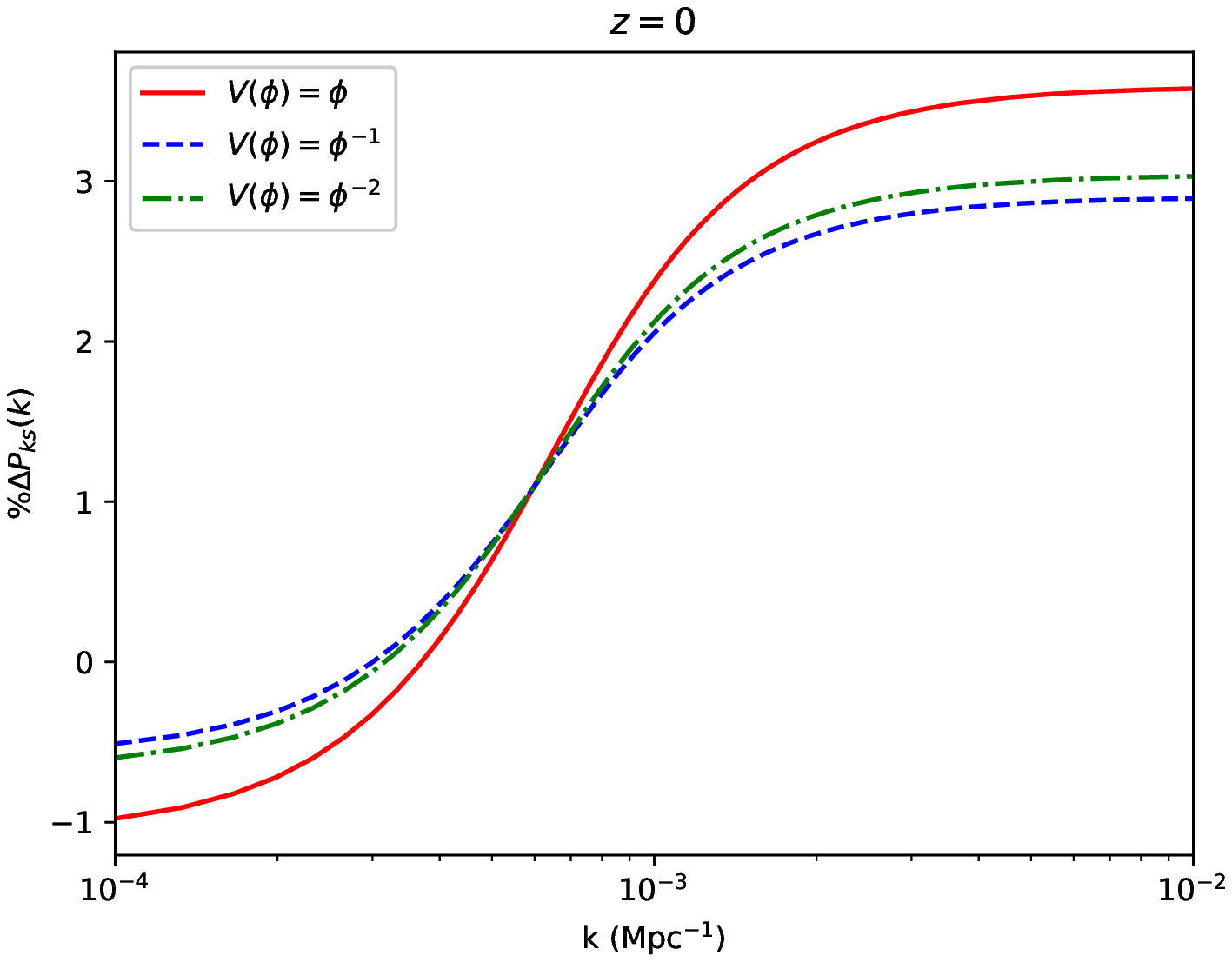}
\includegraphics[width=0.33\linewidth]{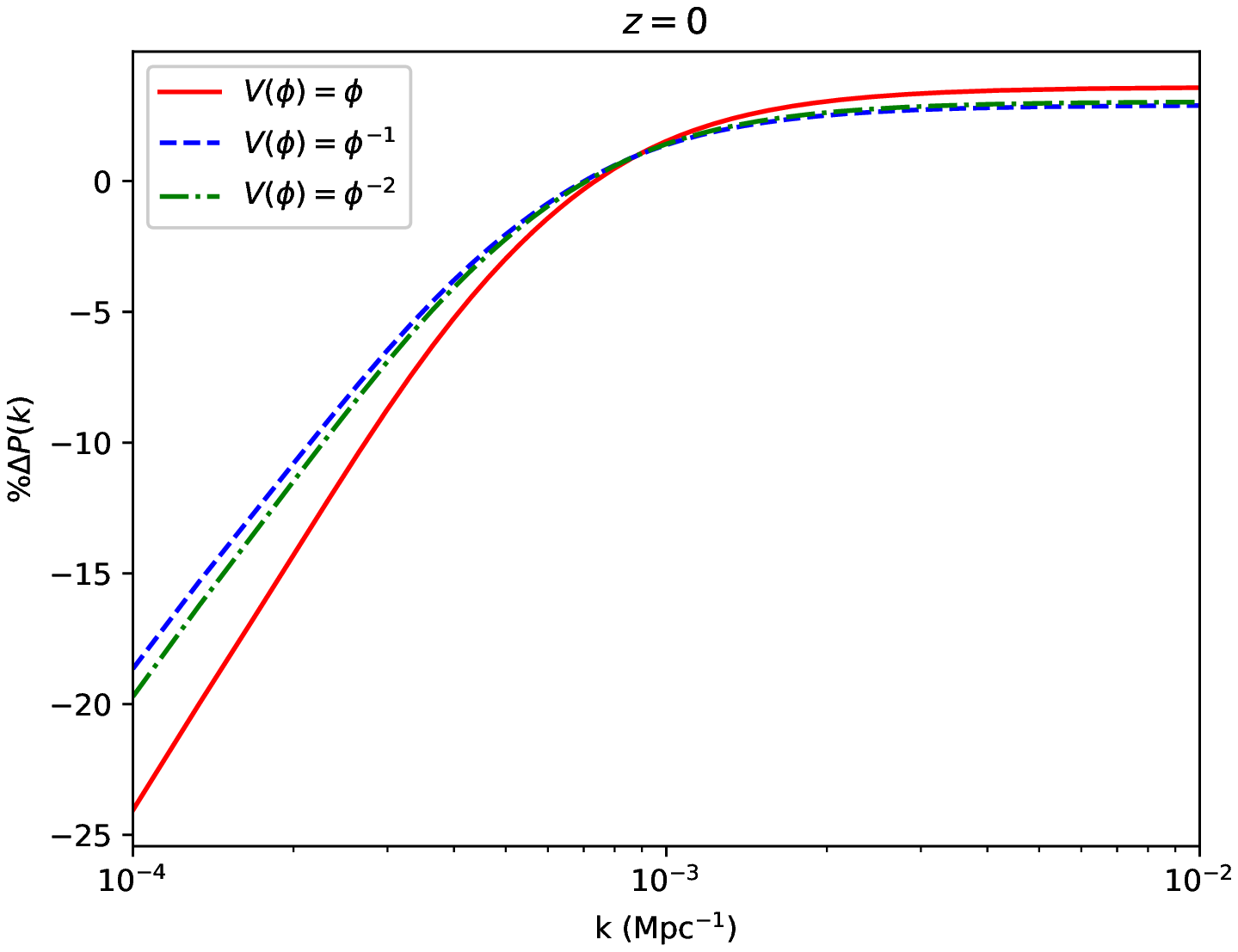}\\
\includegraphics[width=0.33\linewidth]{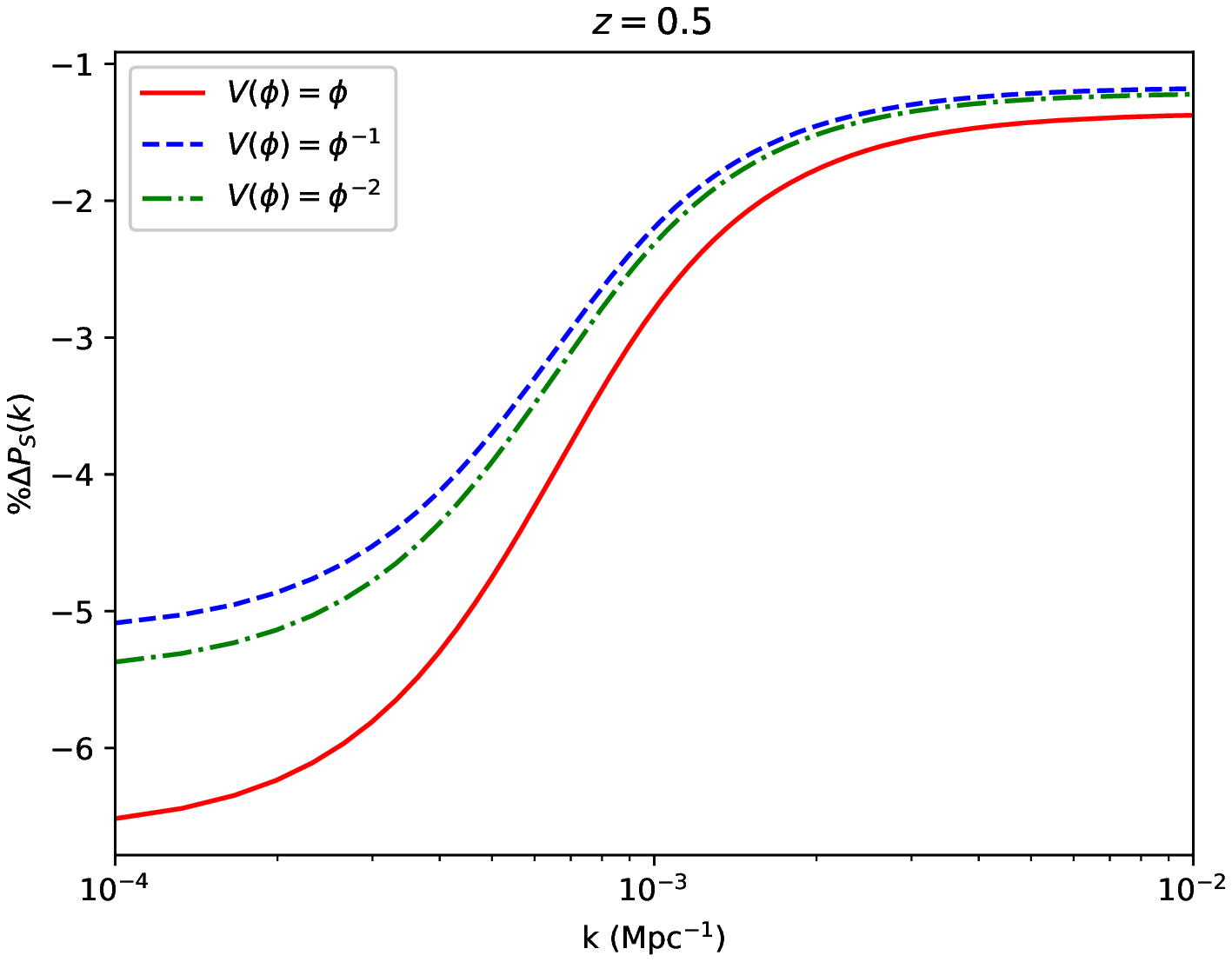}
\includegraphics[width=0.33\linewidth]{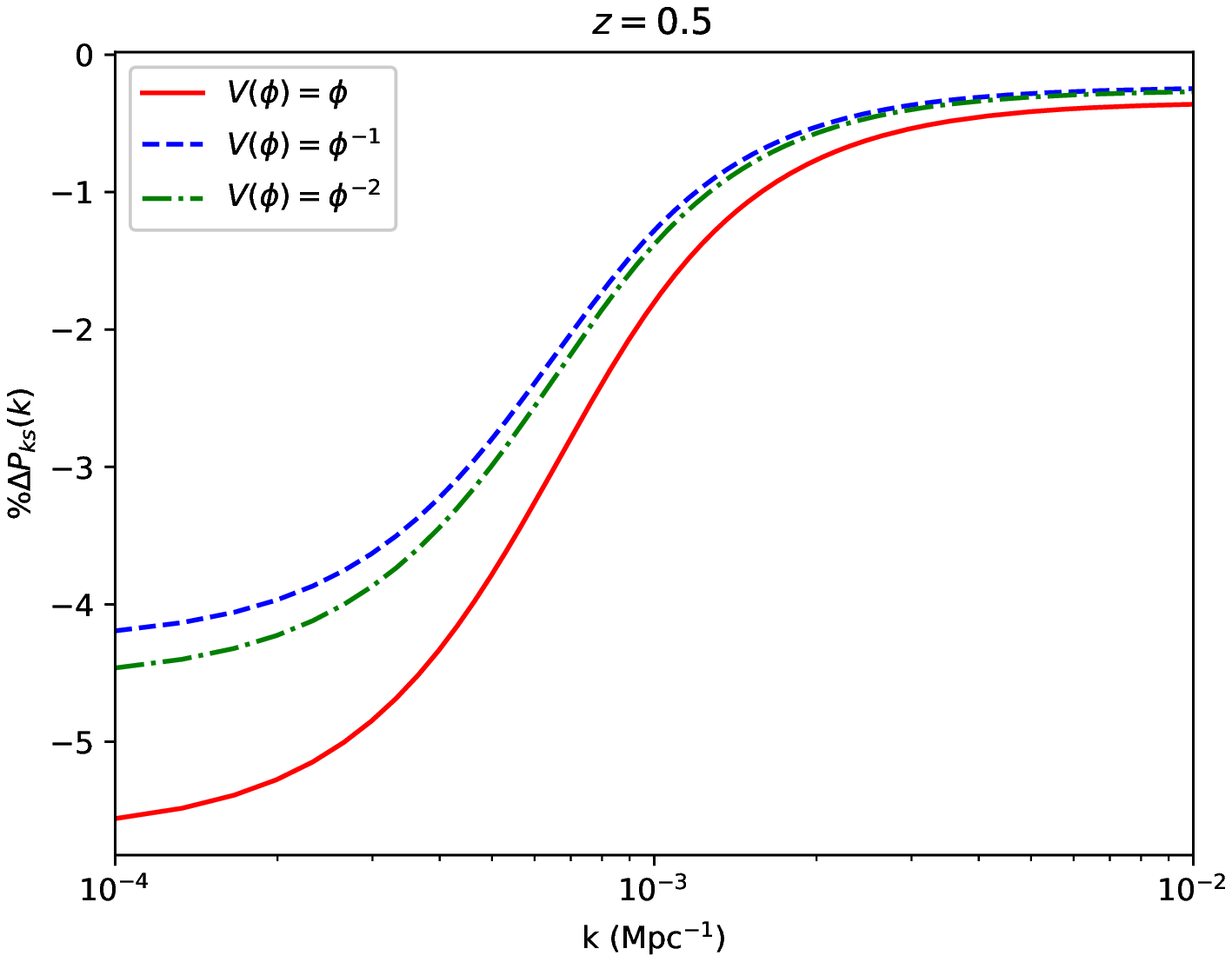}
\includegraphics[width=0.33\linewidth]{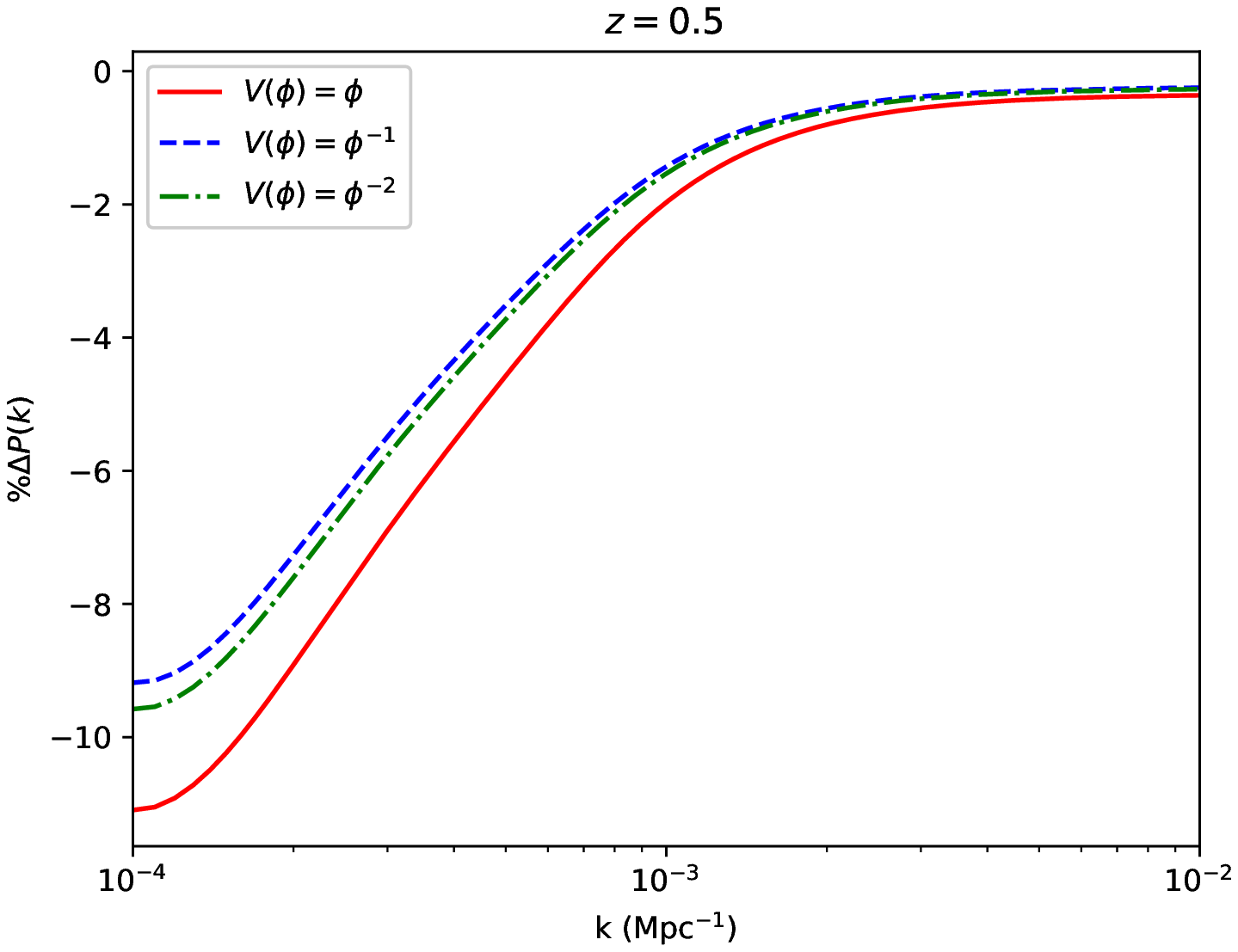}\\
\includegraphics[width=0.33\linewidth]{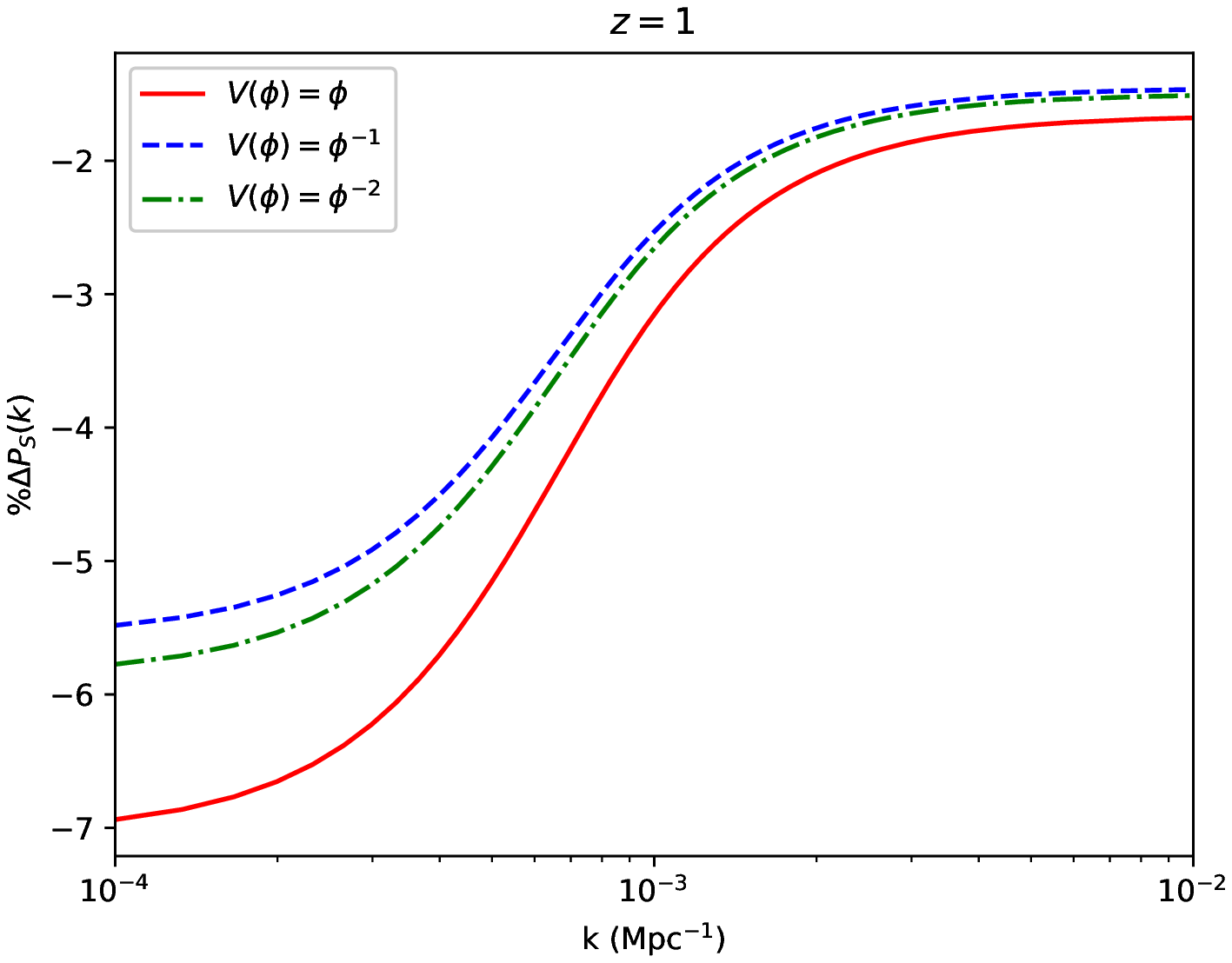}
\includegraphics[width=0.33\linewidth]{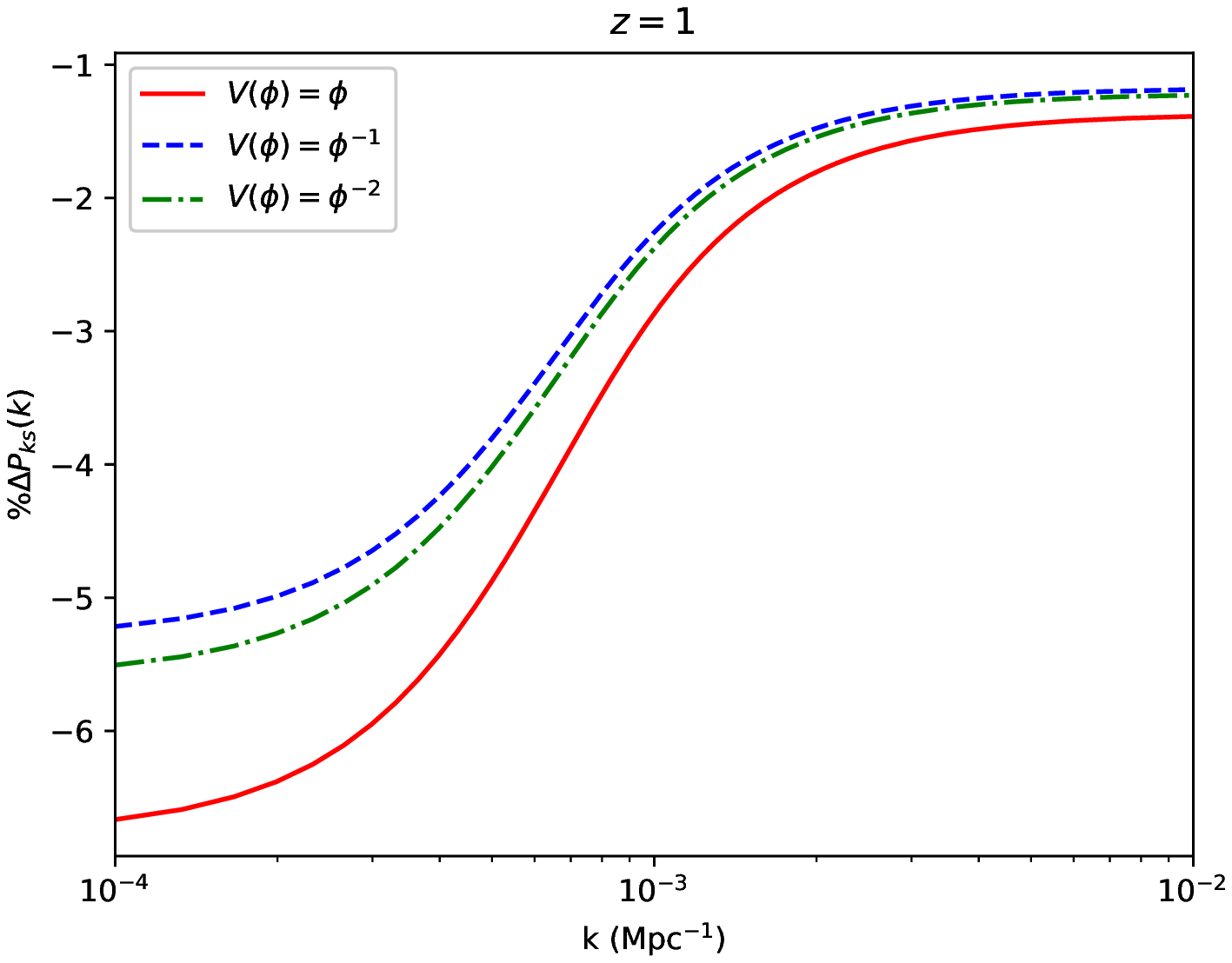}
\includegraphics[width=0.33\linewidth]{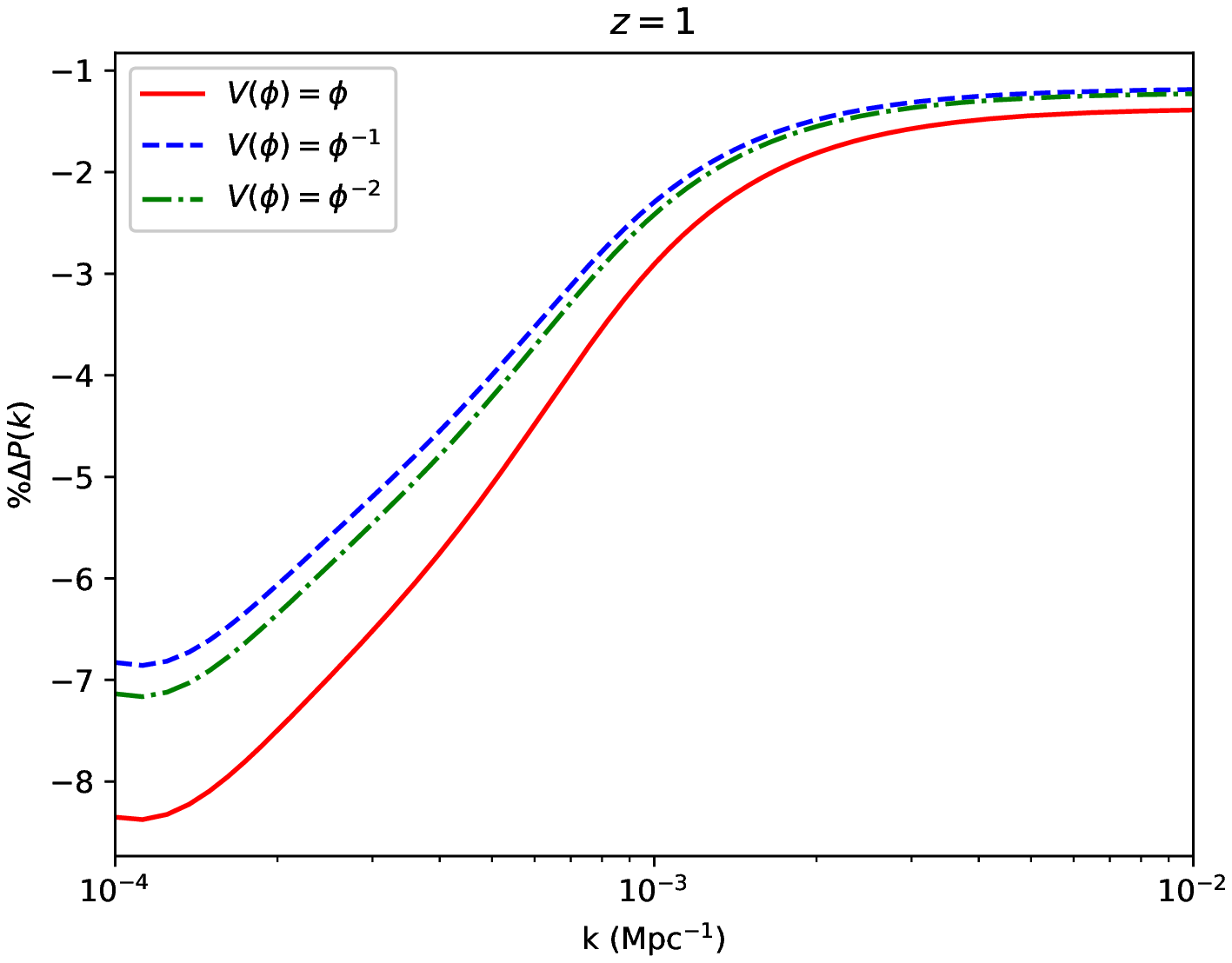}
\end{tabular}
\caption{ Percentage deviation in power spectrum $P(k)$ from $\Lambda CDM$ model for different scalar field potentials, and for different redshifts as a function of $k$. The left column is for the deviation in standard matter power spectrum $P_s(k)$ given by eqn. (\ref{25}), the middle column is for the deviation in the power spectrum with th Kaiser redshift space distortion term $P_{ks}(k)$ and the right column is for the full observed galaxy power spectrum $P(k)$ given by eqn.(\ref{eq24}). }
\label{figf}
\end{figure*}

Considering the different aspects of galaxy distribution, we can study the evolution of our universe. Newtonian perturbations are enough to study the underlying dark matter distribution on sub-horizon scales. On horizon scales, we need full general relativistic treatment to study the effects of dark energy perturbations on dark matter distribution. It will help us to distinguish between different dark energy models from modified gravity models.

We see the galaxies in the redshift space and the galaxy distribution is influenced by the peculiar velocities of the galaxies in addition to the dark matter fluctuations. This gives rise to Kaiser redshift space distortion \cite{Kaiser:1987qv} which is a measure of large scale velocity fields. The gravitational potential in the metric (equation (\ref{eq6})) can affect the photon geodesics by integration along the path and gives rise to the gravitational lensing effect. This effect alters the galaxy distribution and results in magnification bias \cite{Moessner:1997qs}.

In recent past, people have shown that the general relativistic treatment on large scales can affect the observed galaxy distribution by contributing to how the gravitational potential, velocity fields and matter density affect the observed number density of galaxies on large scales \cite{Jeong:2011as,Challinor:2011bk,Bonvin:2014owa,Duniya:2016ibg,Duniya:2015dpa,Yoo:2009au}. These general relativistic effects are negligible on small sub-horizon scales but are significant on large scales and help to distinguish between different dark energy models from modified gravity models.

All the above effects play important part in the observed fluctuations in the number of galaxies across the sky at different redshifts and angles. The galaxy number overdensity $\Delta^o$ incorporating these effects can be written as \cite{Challinor:2011bk,Duniya:2016ibg,Duniya:2013eta,Duniya:2015nva}

\begin{equation}
\Delta^o = \left[{b + f \mu^2} + \mathcal{A} (\frac{\mathcal{H}}{k})^2 +  i\mu\mathcal{B} (\frac{\mathcal{H}}{k})\right]\Delta_{m},
\label{eq21}
\end{equation}

\noindent
where $b$ is the bias parameter on linear scales, $f$ is the redshift space distortion parameter, $\mu = \frac{\vec{n}\cdot\vec{k}}{k}$ with $\vec{n}$ gives the direction of observation, $\vec{k}$ is the wave vector with magnitude $k$. The parameters $\mathcal{A}$ and $\mathcal{B}$, which arise due to full general relativistic treatment, are given by

\begin{equation}
\mathcal{A} = 3f + (\frac{k}{\mathcal{H}})^2 \Big{[} 3 + \frac{\mathcal{H}'}{\mathcal{H}^{2}} + \frac{\Phi'}{\mathcal{H} \Phi} \Big{]} \frac{\Phi}{\Delta_{m}},
\label{eq22}
\end{equation}

\begin{equation}
\mathcal{B} = - \Big{[} 2 + \frac{\mathcal{H}'}{\mathcal{H}^{2}} \Big{]} f.
\label{eq23}
\end{equation}

\noindent
We have assumed constant comoving galaxy number density, thus galaxy evolution bias is zero in our case and we have considered magnification bias $b=1$ \cite{Duniya:2015nva}. We have neglected time-delay, ISW and weak lensing integrated terms in our calculations. In equation (\ref{eq21}), the first term inside the square bracket is related to the galaxy bias, the second term is the Kaiser redshift term, third and fourth terms are purely due to the general relativistic corrections. In the last two terms, $\mathcal{A}$ (given by equation (\ref{eq22})) is related to the peculiar velocity fields (equation (\ref{eq20})) and gravitational potential, and $\mathcal{B}$ is related to the Doppler effect.

We can write the power spectrum for the observed galaxy number overdensity using equation (\ref{eq21}) (only real part) as \cite{Jeong:2011as,Duniya:2013eta}

\ber
\nonumber
P(k,z)= P_{s}(k,z)\Big[(b + f \mu^{2})^{2} + 2 (b + f \mu^{2}) \Big{(} \frac{\mathcal{A}}{x^{2}} \Big{)}\\
 + \frac{\mathcal{A}^{2}}{x^{4}} + \mu^{2} \Big{(} \frac{\mathcal{B}^{2}}{x^{2}} \Big{)}\Big],
\label{eq24}
\eer

\noindent
where $x=\frac{k}{\mathcal{H}}$ and $P_s(k,z)$ is the standard matter power spectrum

\begin{equation}
P_{s}(k,z) = A k^{n_{s}-4} T(k)^{2}\left(\frac{|\Delta_{m}(k,z)|}{|\Phi(k,0)|}\right)^{2}.
\label{25}
\end{equation}

We can also define the power spectrum with only Kaiser redshift space distortion term included as 
\begin{equation}
P_{ks}(k,z) = (b + f \mu^{2})^{2}P_{s}(k,z).
\label{kas}
\end{equation}

In the standard matter power spectrum given by equation (\ref{25}), $A$ is fixed by $\sigma_8$ normalisation. We use the Eisenstein-Hu transfer function $T(k)$ \cite{Eisenstein:1997ik} in our case. In figure (5), we have plotted the line of sight $(\mu=1)$ for the observed galaxy power spectrum at $z=0$ for linear potentials only by Using equation (\ref{eq24}). We put the spectral index for the initial power spectrum $n_s=0.98,\ \sigma_8=0.8,\ \Omega_{bo}=0.05,\ \Omega_{mo}=0.28$ and $h=0.7$ using $\sigma_8$ normalisation.

In figure (5), we have plotted the observed galaxy power spectrum with and without general relativistic corrections. When Kaiser redshift space distortion term is  considered, the power spectrum $P_{ks}(k,z)$ shifts with an almost constant factor to higher values on all scales compared to the standard matter power spectrum. When general relativistic corrections are considered, then total power spectrum remains almost equal with $P_{ks}(k,z)$ on small scales but shows substantial enhancement to higher values on large scales, which again shows that the $GR$ corrections contribute on large scales.

In figure (6), we have shown the percentage deviation in the standard matter power spectrum $P_s(k,z)$, the power spectrum with only Kaiser term $P_{ks}(k,z)$ and the total power spectrum $P(k,z)$ from the $\Lambda CDM$ model on different scales $k$ for different redshifts and for different scalar field potentials. 

From equation (\ref{25}), we can observe how $P_s(k,z)$ depends upon $\Delta_m$ and $\Phi$. From figure (3), except at $z=0$ where there is slight enhancement in $\Delta_m$ in tachyon model compared to $\Lambda$CDM, for other redshifts, the deviation in $\Delta_m$ from $\Lambda$CDM is negligible. On the other hand, the gravitational potential $\Phi$ has a reasonable enhancement in tachyon model compared to $\Lambda$CDM at $z=0$ on large scales. With this, from equation (\ref{25}), one expects the suppressions in $P_s(k,z)$ in tachyon model compared to $\Lambda$CDM on large scales which is shown in the left column in figure (6).

Next we consider the power spectrum with the Kaiser redshift space distortion term $P_{ks}(k,z)$ (equation (\ref{kas})). It depends upon the growth function $f$ given by equation (\ref{eq20}). In figure (4), we have shown that there is an enhancement in $f$ in tachyon model compared to $\Lambda$CDM for large redshift and this enhancement is largely scale dependent. For smaller redshifts, the enhancement is minimal. This reflects the effect in $P_{ks}(k,z)$ (equation (\ref{kas})) as shown in the middle column in figure (6).

Finally we consider the full power spectrum $P(k,z)$ with general relativistic corrections given by equations (\ref{eq22}) and (\ref{eq23}). The deviation in $P(k,z)$ from $\Lambda CDM$ model is large on large scales due to the contribution of dark energy perturbations on large scales and small redshifts through $\Phi$ term (scalar field model starts behaving like matter only model for higher redshifts) . For redshift $z=0$, at large scale the suppression from $\Lambda CDM$ is around $17\%-24\%$ depending upon different scalar field potentials. But if we look at at the smaller scale at $z=0$, we observe a slightly higher $P(k)$ for the present model than the corresponding $\Lambda$CDM. At non-zero redshift, the $P(k)$ remains suppressed even at smaller scale. Comparing this to the deviation in $P_{ks}(k,z)$, we can see that the $GR$ corrections highly suppress the power spectrum at larger scale. At smaller scale, the deviation in $P(k,z)$ has similar behaviour as in $P_{ks}(k,z)$ due to the negligible contribution of the $GR$ corrections on small scales. The effect of GR corrections is maximum around present day. We should also stress that for the full power spectrum $P(k,z)$ with general relativistic corrections, deviation from $\Lambda$CDM in tachyon model is much larger than the corresponding deviations in canonical scalar field models \cite{Dinda:2016ibo} as well as cubic galileon models \cite{Dinda:2017lpz}. This is why tachyon models can be more easily distinguished from $\Lambda$CDM compared to canonical scalar field as well as galileon models.

It worth mention at this point that some earlier work have also emphasized the perturbation in tachyon dark energy and there effects on cosmic large scale structure. Singh, Jassal and Sharma \cite{Singh:2019bfd} have discussed the perturbations in tachyon dark energy and their effects on the clustering of matter. They have studied the evolution of gravitational potential, density contrast of dark matter and dark energy  for inverse square and exponential potential. In the present work, we have adopted linear, inverse and inverse square potential. In \cite{Singh:2019bfd}, it was observed that the effect of dark energy perturbations are significant only at super-horizon scale and it causes the enhancement of gravitational potential and the growth of density contrast at super-horizon scale for tachyon dark energy compared to that of $\Lambda$CDM. This results are totally consistent with the findings of the present work. Besides, in the present work, we have also emphasized on the nature matter power spectrum for tachyon dark energy considering the fully relativistic perturbation equations. Substantial suppression of power is observed at large scale in case of tachyon dark energy. In another recent article, Rajvanshi {et al.} \cite{Rajvanshi:2021afc} have compared the linear perturbations in tachyon and quintessence dark energy and their impacts on the observational measurements of cosmological parameters from cosmic microwave background. It was found that these two models, namely the tachyon and quintessence are not distinguishable at background and linear perturbation level. In the present work, we have shown that the fully relativistic  analysis of the matter and dark energy perturbations enable us to distinguish the present model from $\Lambda$CDM. The study of matter power spectrum using the nonlinear equations of tachyon dark energy perturbations is one of the new aspects of the present study.   Similar analysis for thawing quintessence dark energy has been carried by Dinda and Sen \cite{Dinda:2016ibo}. A close observation of the results from the present analysis and the results in \cite{Dinda:2016ibo} would reveal that though the suppression of power spectrum in case of tachyon and quintessence from the $\Lambda$CDM have similar patterns, the amount of suppressions of power spectrum is not the same in tachyon and quintessence. Thus the comparison of fully relativistic matter power spectrum could successfully break the degeneracy of quintessence and tachyon dark energy cosmology.

\section{Conclusion}
\label{conclu}

The present work deals with the relativistic perturbations in a tachyon field dark energy model. The prime emphasis is on the nature of cosmological perturbations considering the full general relativistic (GR) corrections. The GR corrections are important at large scales where the inhomogeneity in dark energy distribution is no more negligible. We have formed a set of coupled dynamical equations involving the relevant quantities of background and perturbed universe. The solutions of the set of dynamical equations are studied with proper initial conditions. 

The gravitational potential ($\Phi$) is found to be slightly higher than that of $\Lambda CDM$ (figure \ref{figb}). The deviation is higher at large scales, where GR corrections effectively contribute. The deviation is higher at $z=0$. The comoving matter density contrast also shows a similar profile of deviation from the $\Lambda CDM$ (figure \ref{figc}). The linear growth rate of matter perturbation ($f$) is also found to be higher for the present model than the $\Lambda CDM$ and the deviation is maximum at $z=0$ (figure \ref{figd}). Further we have studied power spectrum of matter density contrast and observed galaxy power spectrum for the present model and also investigated the difference in the power spectrum from the $\Lambda CDM$ power spectrum (figure \ref{fige} and \ref{figf}). Suppression in power in the matter power spectrum ($P_s(k)$) compared to the $\Lambda CDM$ is observed and the power suppression is higher at large scale (left column of figure \ref{figf}). The power is enhanced when the Kaiser redshift space distortion term is introduced in the power spectrum ($P_{ks}(k)$) (middle panel of figure \ref{figf}). At large scales, $P_{ks}(k)$ remains suppressed compared to the $\Lambda CDM$ model. But at smaller scales and at $z=0$, the $P_{ks}(k)$ for the present model overtakes the $\Lambda CDM$. At other redshifts, it is almost same as compared to the $\Lambda CDM$ curves at smaller scales. In the right column of figure \ref{figf}, the deviation in observed galaxy power spectrum from $\Lambda CDM$ is shown. The observed galaxy spectrum is also suppressed in the present model at large scales. At smaller scales, it comes closer to the $\Lambda CDM$ spectrum. It is apparent from the plots that the general relativistic corrections, that introduces the effect of dark energy inhomogeneity in cosmological perturbations, suppresses the matter power spectrum and observed galaxy power spectrum substantially at large scales. On the other hand, the power is not much affected by the GR corrections at smaller scales as the dark energy inhomogeneity is not effective at that scales. 

Future observations like SKA, LSST will observe the sky at much larger scale and at much higher redshift. For those observations, GR corrections in the cosmological perturbations are essential. At that scale of observation, the inhomogeneity of dark energy distribution would have its signature on the matter field. Hence those observations will be highly effective to distinguish homogeneous dark energy (the $\Lambda CDM$) from time varying dark energy which allows the clustering of dark energy. Even different time varying dark energy models could be distinguished in this method. Hence this type of studies are highly relevant in present cosmological research. Future observations in radio and optical regime would be highly effective to reveal the nature of dark energy as well as to give a better understanding about the physical entity of the dark energy.

\section*{Acknowledgment}

AB acknowledges the financial support from the Council of Scientific and Industrial Research (CSIR), Government of India as a SRF (CSIR file no. 09/466(0172)/2016-EMR-I).
AM acknowledges the financial support from the Science and Engineering Research Board (SERB), Department of Science and Technology,
Government of India as a National Post-Doctoral Fellow (NPDF file no. PDF/2018/001859). AAS acknowledges funding from DST-SERB, Govt of India, under the project NO. MTR/20l9/000599.


\end{document}